\documentclass[11pt]{article}

\usepackage[final]{acl}

\usepackage{times}
\usepackage{latexsym}

\usepackage[T1]{fontenc}

\usepackage[utf8]{inputenc}

\usepackage{microtype}

\usepackage{inconsolata}

\usepackage{graphicx}
\usepackage{enumitem}
\usepackage{amsmath}
\usepackage{booktabs}
\usepackage{multirow}
\usepackage{etoolbox,caption,subcaption}
\usepackage{listings}
\usepackage{color}
\usepackage{tcolorbox}
\usepackage{pifont}
\usepackage{amssymb}
\usepackage{array}
\usepackage{tabularx}
\usepackage{soul} 
\usepackage{color, xcolor}
\usepackage[linesnumbered,ruled,vlined]{algorithm2e}
\usepackage{cite}
\usepackage{makecell}
\usepackage{colortbl}
\usepackage{bbding}
\usepackage{bm}
\usepackage{dsfont}
\usepackage{comment}

\usepackage{hyperref}
\usepackage{xspace}

\usepackage[utf8]{inputenc}

\newtcolorbox{promptbox}[1]{
  colback=gray!5,      
  colframe=gray!50,    
  coltitle=black,      
  boxrule=0.5pt,       
  sharp corners,       
  fonttitle=\bfseries\small,
  title={#1},
  left=2mm, right=2mm, top=2mm, bottom=2mm, 
  arc=0mm
}

\title{Eliminating Out-of-Domain Recommendations in LLM-based Recommender Systems: A Unified View}

\author{
 \textbf{Hao Liao\textsuperscript{1,}}
 \textbf{Jiwei Zhang\textsuperscript{1}},
 \textbf{Jianxun Lian\textsuperscript{2,}}\thanks{\ \ Corresponding author.},
 \textbf{Wensheng Lu\textsuperscript{1}},
 \textbf{Mingqi Wu\textsuperscript{2,*}},
 \textbf{Shuo Wang\textsuperscript{1}},
 \\
 \textbf{Yong Zhang\textsuperscript{1}},
 \textbf{Yitian Huang\textsuperscript{1}},
 \textbf{Mingyang Zhou\textsuperscript{1,}},
 \textbf{Rui Mao\textsuperscript{1}}
\\
 \textsuperscript{1}College of Computer Science and Software Engineering, Shenzhen University, China
 \\
 \textsuperscript{2}Microsoft Research Asia
\\
\texttt{jianxun.lian@outlook.com}\\
}

\begin{document}
\maketitle

\begin{abstract}
Recommender systems based on Large Language Models (LLMs) are often plagued by hallucinations of out-of-domain (OOD) items. To address this, we propose RecLM, a unified framework that bridges the gap between retrieval and generation by instantiating three grounding paradigms under a single architecture: embedding-based retrieval, constrained generation over rewritten item titles, and discrete item-tokenizer generation. Using the same backbone LLM and prompts, we systematically compare these three views on public benchmarks. RecLM strictly eradicates OOD recommendations (OOD@10 = 0) across all variants, and the constrained generation variants RecLM-cgen and RecLM-token achieve overall state-of-the-art accuracy compared to both strong ID-based and LLM-based baselines. Our unified view provides a systematic basis for comparing three distinct paradigms to reduce item hallucinations, offering a practical framework to facilitate the application of LLMs to recommendation tasks. Source code is at \href{https://github.com/microsoft/RecAI}{https://github.com/microsoft/RecAI}.
\end{abstract}

\section{Introduction}

\begin{figure}[t]
    \centering
    \includegraphics[width=\linewidth]{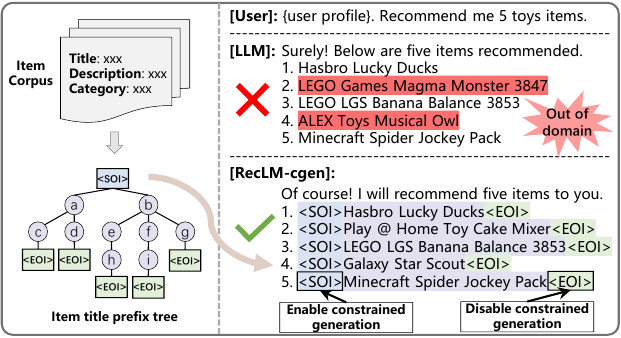}
\caption{An illustration of unconstrained versus constrained decoding.}
    \label{fig:illustration}
\end{figure}

Large language models (LLMs) are increasingly used to build conversational recommender systems, thanks to their strengths in language understanding, reasoning, and instruction following. Prior work either augments LLMs with prompt engineering or agentic retrieval~\citep{yao2023knowledge,gao2023chat,huang2023recommender}, or fine-tunes them with domain knowledge~\citep{DBLP:conf/acl/LuLZL0LX24,10.1145/3708882,10.1007/978-3-031-56063-7_42}, bringing gains in recommendation quality but still suffering from out-of-domain (OOD) item recommendations (as illustrated in Figure~\ref{fig:illustration}) that can harm real-world systems.

Current attempts to mitigate OOD recommendations largely fall into three grounding paradigms. Retrieval-based methods map user and item information into an embedding space and retrieve in-domain items. Constrained generation methods restrict decoding to a catalog-dependent subspace, often via prefix trees over item titles. Item-tokenizer methods instead map each item to a compact sequence of discrete codes and generate over this learned token space. However, these paradigms are typically developed and evaluated in isolation, using different backbones and prompts, which makes it difficult to understand their relative strengths or to flexibly choose between them in practice.

In this work, we address the OOD issue and, more broadly, seek a unified view of these paradigms for LLM-based recommendation. We introduce {\bf \underline{Rec}}ommendation {\bf\underline{L}}anguage {\bf \underline{M}}odels ({\bf RecLM}), a unified framework that bridges the gap between retrieval and generation by instantiating three complementary grounding strategies under a single architecture and training protocol. The key idea is to teach the LLM to first emit a special start-of-item token \texttt{<SOI>} to mark where recommendations should appear, and then plug in different grounding mechanisms after \texttt{<SOI>} while keeping the rest of the conversational behavior unchanged, as illustrated in Figure~\ref{fig:illustration}. Concretely, RecLM-ret uses the hidden state at \texttt{<SOI>} to retrieve an in-domain item from an embedding index, RecLM-cgen generates a rewritten item title under a prefix-tree constraint built over RL-optimized titles, and RecLM-token generates a short sequence of learned item tokens under a prefix tree and maps it back to a catalog item.   

Our experiments on three public datasets demonstrate that all three {RecLM} variants successfully eliminate out-of-domain recommendations (OOD@10 = 0). Moreover, the constrained generation variants ({RecLM-cgen} and {RecLM-token}) achieve state-of-the-art accuracy, outperforming both strong ID-based and LLM-based baselines. By integrating retrieval, constrained generation, and item tokenization into a unified framework, {RecLM} enables a reliable comparison of these distinct paradigms and offers best practices for designing LLM-based recommendation systems in both academic and industrial settings.


To summarize, our main contributions are:

\begin{itemize} [leftmargin=*]
\item We propose \textsc{RecLM}, a unified framework that trains an LLM to first emit a special start-of-item token (\texttt{<SOI>}) and then delegate recommendation to interchangeable grounding modules. This design unifies retrieval-based and generation-based recommendation paradigms within a single architecture and evaluation protocol.

\item Within this framework, we introduce lightweight enhancement modules, including a reinforcement learning–based Title Rewriter and scope-mask training. These components compress verbose item metadata into concise, human-readable identifiers and improve the alignment between constrained generation and recommendation objectives under limited token budgets.

\item Experimental results show that the three paradigms exhibit clear differences in ranking accuracy, OOD rate, efficiency, and conversational behavior across three public recommendation datasets. All RecLM variants strictly eliminate OOD recommendations (OOD@10 = 0), and the observed trade-offs reveal complementary strengths across retrieval, constrained textual generation, and item tokenization, offering practical guidance for method selection in different deployment scenarios.
\end{itemize}

\section{Related Work}

\subsection{LLMs for Recommender Systems}  

LLMs have significantly influenced various NLP applications, including recommender systems. 
Their potential has been widely recognized in facilitating a new type of generative recommender systems~\citep{wu2024survey,lian2024recai,lyu2023llm,10.1007/978-3-031-56063-7_42}. 
\citet{said2025explaining} provides a comprehensive review of the literature on using LLMs for generating recommendation explanations. 
Methods for selectively injecting domain-specific knowledge into prompts to enhance the recommendation capabilities of LLMs without fine-tuning are introduced by \citet{yao2023knowledge} and \citet{bacciu2024generating}.   
Another line of research focuses on fine-tuning LLMs to inject domain knowledge, demonstrating significant improvements in recommendation performance~\citep{10.1145/3708882,DBLP:conf/acl/LuLZL0LX24,yang2023palr,zhu2024collaborative}. 
However, these approaches often face the challenge of OOD item generation, where LLMs may recommend items that are not present within the current domain, potentially leading to negative business impacts.  
   
\subsection{Addressing Out-of-domain Recommendations}  

The issue of OOD item generation is a critical challenge in LLM-based recommenders.  
\citet{bao2023bi} proposes a generate-then-align method to ensure that recommended items are grounded within the domain item set. 
\citet{gao2023chat} and \citet{huang2023recommender} leverage agentic frameworks where LLMs act as controllers and natural language interfaces for user interactions. 
When making recommendations, these frameworks call traditional recommender models to retrieve relevant items.   
Another promising direction is constrained generation.
This paradigm restricts the LLM's decoding space to a subspace conditioned by the context, thereby avoiding OOD generation~\citep{dong2024xgrammar}. Constrained generation methods maintain the traditional language generation process without necessitating significant modifications to the LLM. In addition to retrieval-based grounding and prompt-based constrained generation, recent work proposes an item-tokenizer paradigm for LLM-based recommenders, where each item is mapped to a compact sequence of discrete tokens and decoded under prefix-tree constraints~\citep{10.1145/3626772.3657821,10.1145/3627673.3679569,zheng2024adapting,lin2025order}. These methods tightly couple catalog structure with generative modeling, but are typically studied in isolation from retrieval- and text-based grounding. From a unified perspective, our work brings together these three lines. It allows us to interpret existing approaches as points in a common design space and to empirically compare their behavior for OOD mitigation under the same backbone model and evaluation protocol.

\section{Methodology}

Our goal is to avoid recommending OOD items while preserving the conversational strengths of LLMs. To this end, we build three RecLM variants under two fundamental paradigms—in-domain retrieval and constrained catalog grounding—and implement them within a single lightweight framework that introduces minimal changes to the backbone model. The overall fraemwork is illustrated in Figure~\ref{fig:model_overview}. The key mechanism is a pair of special item indicator tokens that tell the LLM when it is entering and leaving a recommendation segment, so that different grounding strategies can be plugged into the same decoding process.

\subsection{Special Item Indicator Token} 
\label{sec:control_symble}   

\begin{figure*}[ht]
    \centering
    \includegraphics[width=1.0\textwidth]{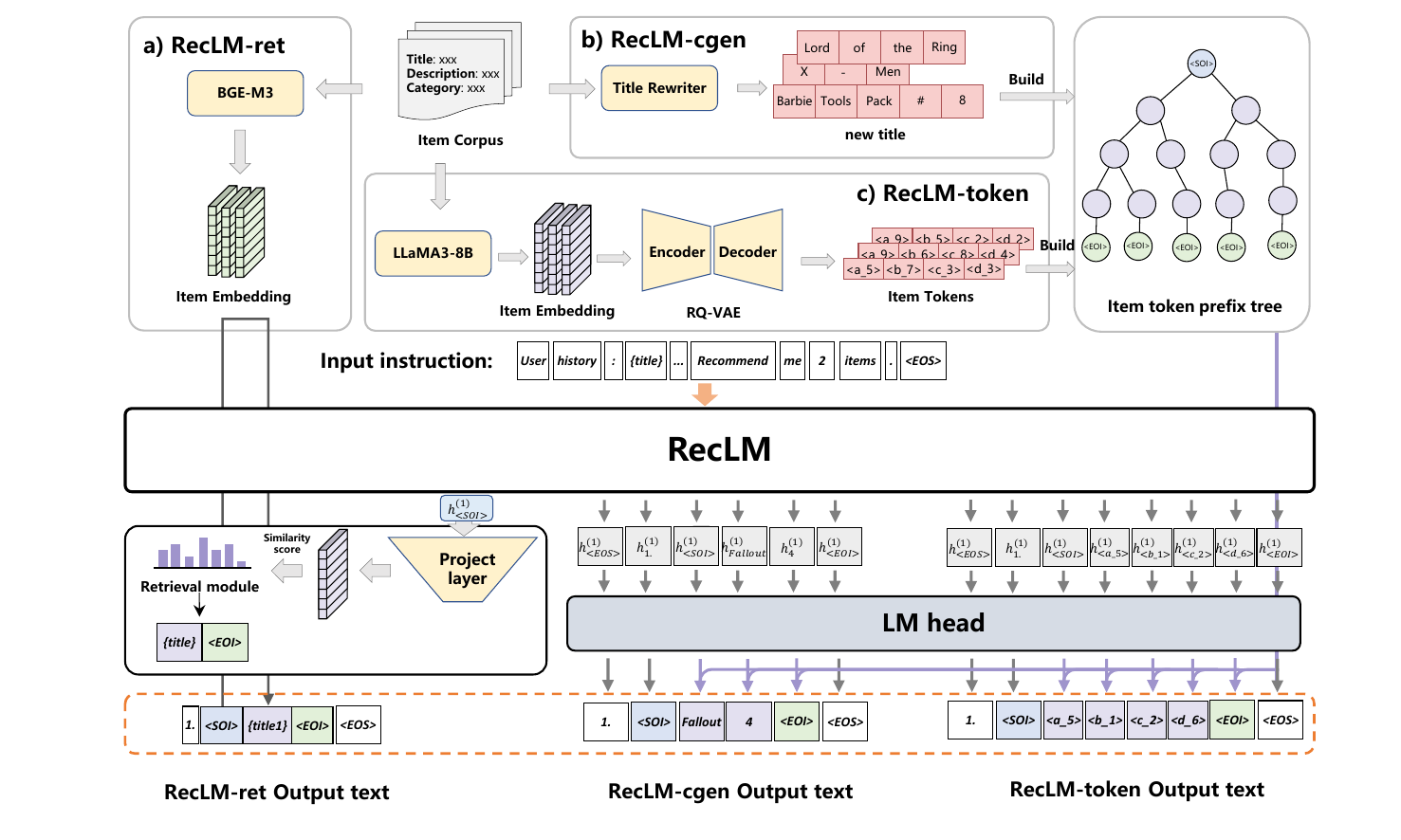}
    \caption{Overview of RecLM variants: embedding-based retrieval (RecLM-ret), constrained generation over rewritten item titles (RecLM-cgen), and discrete item-tokenizer generation (RecLM-token).}
    \label{fig:model_overview}
\end{figure*}

We equip the backbone LLM with two special tokens—\texttt{<SOI>} (start-of-item) and \texttt{<EOI>} (end-of-item)—to explicitly mark recommendation segments in its outputs. After fine-tuning on recommendation data, RecLM learns to produce sequences of the form \emph{\texttt{<SOI>} item identifier \texttt{<EOI>}} at appropriate positions in a conversation. The emission of \texttt{<SOI>} signals that the model is entering an item segment where grounding constraints apply, and the appearance of \texttt{<EOI>} token marks its termination, after which the model resumes generating general text.  As illustrated in Figure~\ref{fig:model_overview}, what happens between \texttt{<SOI>} and \texttt{<EOI>} is then delegated to one of the three RecLM variants: retrieval over an item index (RecLM-ret), constrained decoding over rewritten titles (RecLM-cgen), or constrained decoding over discrete item tokens (RecLM-token).

\subsection{RecLM-ret}

RecLM-ret instantiates retrieval-style grounding within our unified framework. We build a domain-specific item index by encoding each item's title, description, and category with BGE-M3~\citep{chen-etal-2024-m3}, followed by a lightweight adapter that maps embeddings to the target space, yielding $\mathcal{E} = \{\mathbf{e}_i\}$. At inference time, when RecLM emits \texttt{<SOI>}, the corresponding hidden state $\mathbf{h}_{<SOI>}$ is projected into the item embedding space and the nearest item in $\mathcal{E}$ is retrieved; its title is then inserted into the output and closed by \texttt{<EOI>}. This design reuses standard embedding-based retrieval while cleanly fitting the shared \texttt{<SOI>}/\texttt{<EOI>} interface.

For training, sequence data <$I^{(1...n)}_{history}$, $I^{(1...K)}_{rec}$> are converted into instruction–response pairs <Instruction:$X$, Response:$Y$>, where $X$ encodes the user history and $Y$ contains the recommended items; we follow \citet{DBLP:conf/acl/LuLZL0LX24} for data augmentation and provide prompt templates in Appendix~\ref{app:Prompts}. RecLM-ret is optimized with a language-modeling loss over non-item tokens and an auxiliary retrieval loss that aligns each \texttt{<SOI>} hidden state with the embedding of its ground-truth item:

\begin{equation}\label{eq:loss_lm_masked}
\mathcal{L}_{\text{lm}} = \sum_{\substack{j = 1 \\Y_j\notin \text{\{item, \texttt{<EOI>}\}}}}^{|Y|} -\log P_{\theta}(Y_j|Y_{<j}, X)
\end{equation}

A retrieval loss further teaches the model to select the correct item in the embedding space. Let $\mathbf{h}_{<SOI>}^{(1...K)}$ be the hidden states at \texttt{<SOI>} for the $K$ recommended items in $Y$, and $\text{proj}_{\phi}$ a projection layer. We match the projected vectors to their ground-truth item embeddings in $\mathcal{E}$ using:
\begin{equation}\label{eq:loss_ret}
\mathcal{L}_{\text{ret}} = -\frac{1}{K} \sum_{j=1}^{K}  \log(\sigma(\text{proj}_{\phi}(\mathbf{h}^{(j)}_{<SOI>})\cdot \mathbf{e}_{j}))
\end{equation}
\begin{equation}\label{eq:loss_reclm_ret}
\mathcal{L}_{\text{RecLM-ret}} = \mathcal{L}_{\text{lm}} + \alpha_{\text{ret}} * \mathcal{L}_{\text{ret}}
\end{equation}
where $\alpha_{\text{ret}}$ balances conversational modeling and retrieval alignment. This design keeps the conversational behavior of the backbone LLM largely intact while providing a simple, index-based grounding mechanism compatible with the \texttt{<SOI>}/\texttt{<EOI>} interface.

\subsection{RecLM-cgen}

RecLM-cgen instantiates text-based constrained grounding with rewritten titles. It first employs a Title Rewriter (TR) module to transform each item's verbose title and description into a new, concise, human-readable title. These rewritten titles are used both to construct user histories and to build a prefix tree over the catalog. During generation, once RecLM emits \texttt{<SOI>}, decoding is restricted to paths in the prefix tree until \texttt{<EOI>} appears, ensuring that all recommended items come from the catalog while still leveraging the LLM's language modeling capacity around the title.

\subsubsection{Title Rewriter}

Item IDs are often lengthy or uninformative, whereas descriptions contain richer semantics but are too verbose under token budget constraints.
The TR module addresses this by transforming raw item metadata into compact, human-readable titles that retain key semantic information and serve as the surface forms used in constrained generation.

Since title rewriting is inherently open-ended with no single ground-truth title, we train TR using the Group Relative Policy Optimization (GRPO) algorithm~\citep{deepseek-math}. 
We design five reward components that reflect our design goals: item-to-item similarity (I2I), user-to-item alignment (U2I), decoding complexity (DC), conciseness (CR), and discriminative power (DPR).
These components are combined into a single reward (details in Appendix \ref{reward_design}):

\begin{equation}
\label{eq:reward}
\begin{aligned}
R = {} & \lambda_{1} R_{\text{U2I}} + \lambda_{2} R_{\text{I2I}} + \lambda_{3} R_{\text{DC}} \\
       & + \lambda_{4} R_{\text{CR}} + \lambda_{5} R_{\text{DPR}}.
\end{aligned}
\end{equation}

\paragraph{Recommendation-oriented Reward.}
To enhance recommendation performance, we consider both item-to-item similarity and user-to-item alignment. From the \emph{\bf item-to-item} perspective, the generated short titles should preserve the neighborhood structure in the item space. We construct a contribution matrix from the original data as ground-truth item similarity, and define:
\begin{equation}
R_{\text{I2I}} = 0.5\bigl(1+\text{Spearman}(\pi_{\text{orig}}, \pi_{\text{gen}})\bigr)
\end{equation}
where $\pi_{\text{orig}}$ represents the ground-truth ranking of similar items, derived from the contribution matrix, and $\pi_{\text{gen}}$ denotes the ranking based on cosine similarities between the embedding of the rewritten item title and those of all other items.

From the perspective of \emph{\bf user-to-item}, the objective is to evaluate whether the generated titles better reflect user preferences and improve retrieval of the target item. User-to-Item Reward ($R_{\text{U2I}}$), applied to group-level tasks where TR rewrites a set of item names, measures how well the rewritten prompt preserves the target item's rank. Given the similarity ranking of the ground-truth item $i^*$, we define:
\begin{equation}
    R_{\text{U2I}} = \exp\bigl(-(\text{rank}(i^*)-1)/\tau\bigr), \ \tau=2000
\end{equation}

\paragraph{Decoding Complexity.}
To keep titles easy for LLMs to process, we assess their decoding complexity using conditional perplexity and define:
\begin{align}
    R_{\text{DC}}=\exp(-\alpha_{\text{ppl}} \cdot \text{PPL}(y|X)) 
\end{align}
where lower conditional perplexity corresponds to higher reward.

\paragraph{Conciseness.}
To encourage brevity, we introduce a length-based reward that favors shorter rewritten titles:
\begin{equation}
    R_{\text{CR}} = \bigl(1 + (|y|/|x|)^2\bigr)^{-1}
\end{equation}
where, $|x|$ and $|y|$ denote the number of tokens of the original and generated titles, respectively.

\paragraph{Discriminative Power.}
Finally, rewritten titles should be distinguishable from the titles of semantically similar items. We design a discrimination task where the generated title is used as a prompt to a language model, which is then asked to identify the correct original title from a set of four candidates: the true original title and the titles of the three most similar items. The reward function is:
\begin{equation}
    R_{\text{DPR}}=\mathbb{I}[\text{correct}]
\end{equation}
where, $\mathbb{I}[\text{correct}]$ is an indicator function that returns 1 if the model selects the correct title, and 0 otherwise.

\subsubsection{Scope Mask Training}

During constrained decoding, the next token is chosen from a prefix-tree–defined subset rather than the full vocabulary.
To match this behavior at training time, we introduce a scope mask loss for RecLM-cgen: when computing the loss for item-title tokens, only tokens allowed by the prefix tree are included in the $softmax$ denominator:
\begin{equation}\label{eq:loss_lm_domain}
\mathcal{L}_{\text{cgen}}^{\text{sm}} = \sum_{j=1}^{|Y|} -\log \frac{\exp(\text{logit}(Y_j|Y_{<j},X,\theta))}{\sum\limits_{t\in \text{NT}(Y_{<j})}\exp(\text{logit}(t|Y_{<j},X,\theta))}
\end{equation}

Here, $\text{NT}(Y_{<j})$ returns the valid next-token set given the current prefix.
For general text (e.g., after \texttt{<EOI>}), it equals the full vocabulary; for item titles (between \texttt{<SOI>} and \texttt{<EOI>}), it is restricted to tokens available in the catalog prefix tree.

\subsection{RecLM-token}

RecLM-token casts recommendation as sequence generation over a finite vocabulary of discrete item tokens rather than natural-language titles: each catalog item is assigned a short code sequence (e.g., \verb|<a_11><b_2><c_135><d_157>|), generated between \texttt{<SOI>} and \texttt{<EOI>} and deterministically mapped back to an item.
To construct these codes, we encode item text with Llama3-8B-Instruct to obtain semantic embeddings, discretize them into short code sequences via an RQ-VAE codebook ~\citep{lee2022autoregressive}, and then fine-tune (plus RL) the LLM on tokenized recommendation sequences.

\subsubsection{Item Tokenizer}

Given the semantic embedding $\mathbf{e}_{sem} \in \mathbb{R}^{d_{\text{Llama3-8B}}}$ of each item, a multi-layer perceptron projects it into a latent vector suitable for quantization:
\begin{equation}
    \mathbf{z}_e = \text{EncoderMLP}(\mathbf{e}_{sem}),
\end{equation}
where $\mathbf{z}_e$ serves as the continuous representation to be discretized.

We then apply residual vector quantization (RQ) to turn $\mathbf{z}_e$ into a short sequence of discrete codes.
Given quantization depth $D$, at each stage $d$, we select a code vector $\mathbf{z}_{q}^{(d)}$ from codebook $\mathcal{C}_d$ that best matches the current residual, obtaining the final quantized representation:
\begin{equation}
    \hat{\mathbf{z}}_q = \sum_{d=1}^{D} \mathbf{z}_{q}^{(d)}.
\end{equation}
The corresponding index tuple $t_i = (k_1, \dots, k_D)$ serves as the item's tokenized identifier. Finally, a decoder network $\text{DecoderMLP}$ reconstructs the semantic embedding $\hat{\mathbf{e}}_{sem}$ from $\hat{\mathbf{z}}_q$.

The tokenizer is trained end-to-end with a loss that balances reconstruction fidelity and quantization quality:
\begin{align}
    \mathcal{L}_{total} &= \| \hat{\mathbf{e}}_{sem} - \mathbf{e}_{sem} \|_2^2 + \mathcal{L}_{quant}, \\
    \mathcal{L}_{\text{quant}} &= \sum_{d=1}^{D} \ell_{\text{VQ}}(\mathbf{r}_{d-1}, \mathbf{z}_q^{(d)}),
\end{align}
where $\ell_{\text{VQ}}(\mathbf{r}, \mathbf{z}) = \| \text{sg}[\mathbf{r}] - \mathbf{z} \|_2^2 + \beta \| \mathbf{r} - \text{sg}[\mathbf{z}] \|_2^2$. Here, $\mathbf{r}_{d-1}$ denotes the residual at depth $d$ (with $\mathbf{r}_0 = \mathbf{z}_e$), and $\text{sg}[\cdot]$ is the stop-gradient operator.

The reconstruction loss encourages faithful recovery of the semantic representation. The quantization loss $\mathcal{L}_{quant}$ follows standard residual quantization formulations and includes a commitment term weighted by $\beta$; in all experiments, we set $\beta=0.25$.

\subsubsection{Reinforcement Learning for RecLM-token}

After obtaining discrete item codebook, we first perform supervised fine-tuning to align the language model with the codebook.
Unlike RecLM-cgen, the recommendation segments are expressed purely in terms of discrete identifiers (e.g., \verb|<a_128>|), providing a fully symbolic interface between tokenizer and generator.
We then apply GRPO to further refine generation behavior.

\paragraph{Reward Design.}
We optimize RecLM-token with a reward defined over ranked lists, decomposed into position-sensitive and inclusion-based components. When the target item appears in the generated list, we assign a reward that decreases with its rank position:
\begin{equation}
R_{\text{ord}} =
\begin{cases}
\frac{1}{\log_2(\text{rank} + 1)}, & \text{if the target appears} \\
0, & \text{otherwise}
\end{cases}
\end{equation}
where $\text{rank}$ denotes the 1-based index of the target item.

To complement position-sensitive feedback, we add a reward that depends only on whether the target item appears in the prediction:
\begin{equation}
R_{\text{pre}} = \mathbb{I}(\text{target} \in \text{prediction}),
\end{equation}
capturing coarse-grained relevance at the list level.


\subsection{Multi-round Dialogue Training}

To enable the LLM-based recommendation system to interact naturally with users beyond single-turn question-answering, we incorporate multi-round dialogue training. Without this, training solely on single-turn SFT samples \texttt{<Instruction: $X$, Response: $Y$>} biases the model toward a QA-style recommendation tool and weakens its conversational ability. We therefore augment about 10\% of the data with multi-round conversation (MRC) samples, built by combining a randomly sampled ShareGPT dialogue\footnote{\url{https://huggingface.co/datasets/anon8231489123/ShareGPT_Vicuna_unfiltered}} with a single-turn recommendation task. To diversify contexts, the recommendation turn appears before or after the dialogue with equal probability.

\section{Experiments} 
\subsection{Experiment Settings}

We conduct experiments on three public sequential recommendation datasets: \textbf{Steam}\footnote{\url{https://www.kaggle.com/datasets/antonkozyriev/game-recommendations-on-steam}}, Amazon Movies \& TV\footnote{\url{https://mcauleylab.ucsd.edu/public_datasets/data/amazon_v2/categoryFiles}\label{fn:dataurlamazon}} (\textbf{Movies}), and Amazon Toys \& Games\footref{fn:dataurlamazon} (\textbf{Toys})~\citep{ni2019justifying}. Users with fewer than 17 interactions are filtered out, and each interaction sequence is truncated to the 17 most recent items. After this, 10k users are randomly sampled for experiments.
Following prior work~\citep{kang2018self,DBLP:conf/acl/LuLZL0LX24}, we adopt a leave-one-out protocol: 
for each user, the last interaction is reserved for testing, the second-to-last for validation, and the remaining 15 for training; dataset statistics are summarized in Table~\ref{tab:DatasetStatistic}.

\paragraph{Backbone and Fine-tuning.}
We use Llama3-8B-Instruct as the backbone for all RecLM variants.User histories are truncated to 10 interactions and embedded into instruction-style prompts; the maximum input and output lengths are both 512 tokens.
All models are fine-tuned with LoRA on all linear layers using Adam (learning rate $1\times10^{-4}$, LoRA rank $r=16$, scaling factor $\alpha=8$, batch size $2$), typically converging within 20 epochs.
The token embeddings of \texttt{<SOI>} and \texttt{<EOI>} are initialized as the average embeddings of the phrases "start of an item" and "end of an item".
All reported results are averaged over five runs, with significance assessed using paired tests ($p<0.05$).

In the RecLM-token, we employ residual vector quantization with four codebooks of 256 entries each, yielding a four-token discrete representation per item. During RL training stage, we sample 16 candidate recommendation lists per prompt for relative reward normalization, generate with temperature $1.0$ and maximum length 128, and optimize with learning rate $1\times10^{-5}$, batch size 32. The LoRA rank is 16. Detailed experimental settings can be found in Appendix \ref{appendix:another_traing_details}.

\subsubsection{Metrics}

We evaluate recommendation accuracy with Top-$k$ Hit Ratio ($HR@k$) and Top-$k$ Normalized Discounted Cumulative Gain ($NDCG@k$).
To assess reliability, we additionally report $Repeat@k$, the fraction of duplicate items within the Top-$k$ list, and $OOD@k$, the fraction of Top-$k$ items that fall outside the domain catalog.

\newcolumntype{a}{>{\columncolor{gray!20}}c}

\begin{table*}[]
\setlength{\abovecaptionskip}{0.1cm}
\centering
\resizebox{0.99\textwidth}{!}{
\begin{tabular}{l|aa|ccc|ccc|cccc|ccc}
\toprule
& \multicolumn{2}{a|}{Traditional Recommenders} & \multicolumn{3}{c|}{LLMs (frozen)} & \multicolumn{3}{c|}{LLMs (finetuned)} & \multicolumn{4}{c|}{LLMs (item tokenizer)} & \multicolumn{3}{c}{LLMs (ours)} \\
\cmidrule{2-16}
Metrics & {SASRec} & {GRU4Rec} & {GPT-4o} & {Llama3} & {Llama3-cgen} & {BIGRec} & {CtrlRec} & {PALR} & {IDGenRec} & {SETRec} & {LC-Rec} & {LETTER} & RecLM-ret & RecLM-cgen & RecLM-token \\
\hline
\multicolumn{16}{c}{Dataset: Steam} \\
\hline

$\mathrm{HR@10 \uparrow}$ & $0.0694$ & $0.0599$ & $0.0383$ & $0.0230$ & $0.0261$ & $0.0396$ & $\underline{0.0756}$ & $0.0739$ & $0.0682$ & $0.0626$ & $0.0725$ & $0.0569$ & $0.0600$ & $\textbf{0.0868(+14.8\%)}$ & $0.0733$\\
$\mathrm{NDCG@10 \uparrow}$ & $0.0308$ & $0.0281$ & $0.0194$ & $0.0120$ & $0.0125$ & $0.0244$ & $0.0367$ & $\underline{0.0408}$ & $0.0344$ & $0.0303$ & $0.0390$ & $0.0287$ & $0.0291$ & $\textbf{0.0456(+11.8\%)}$ & $0.0388$\\
$\mathrm{HR@5 \uparrow}$ & $0.0428$ & $0.0323$ & $0.0234$ & $0.0136$ & $0.0147$ & $0.0291$ & $0.0507$ & $0.0488$ & $0.0416$ & $0.0346$ & $0.0473$ & $0.0335$ & $0.0359$ & $\textbf{0.0579(+2.1\%)}$ & $\underline{0.0567}$\\
$\mathrm{NDCG@5 \uparrow}$ & $0.0224$ & $0.0193$ & $0.0147$ & $0.0090$ & $0.0088$ & $0.0201$ & $0.0318$ & $0.0305$ & $0.0248$ & $0.0213$ & $0.0309$ &$0.0212$ & $0.0214$ & $\textbf{0.0361(+9.1\%)}$ & $\underline{0.0331}$\\
$\mathrm{repeat@10 \downarrow}$ & $-$ & $-$ & $1.07\%$ & $2.06\%$ & $\textbf{0.00\%}$ & $\textbf{0.00\%}$ & $1.08\%$ & $1.05\%$ & $\textbf{0.00\%}$ & $\textbf{0.00\%}$ & $\textbf{0.00\%}$ & $\textbf{0.00\%}$ & $\textbf{0.00\%}$ & $\textbf{0.00\%}$ & $\textbf{0.00\%}$\\
$\mathrm{OOD@10 \downarrow}$ & $-$ & $-$ & $16.08\%$ & $15.26\%$ & $2.59\%$ & $\textbf{0.00\%}$ & $2.40\%$ & $2.46\%$ & $\textbf{0.00\%}$ & $\textbf{0.00\%}$ & $\textbf{0.00\%}$ & $\textbf{0.00\%}$ & $\textbf{0.00\%}$ & $\textbf{0.00\%}$ & $\textbf{0.00\%}$\\
\hline
\multicolumn{16}{c}{Dataset: Movies} \\
\hline

$\mathrm{HR@10 \uparrow}$ & $0.1510$ & $0.0722$ & $0.0046$ & $0.0049$ & $0.0246$ & $0.0861$ & $0.1347$ & $0.1335$ & $0.1243$ & $0.0929$ & $\underline{0.1607}$ & $0.1205$ & $0.1145$ & $0.1467$ & $\textbf{0.1700(+5.8\%)}$\\
$\mathrm{NDCG@10 \uparrow}$ & $0.1351$ & $0.0556$ & $0.0028$ & $0.0025$ & $0.0106$ & $0.0760$ & $0.1248$ & $0.1244$ & $0.1064$ & $0.0823$ & $\underline{0.1458}$ & $0.1097$ & $0.1052$ & $0.1311$ & $\textbf{0.1622(+11.2\%)}$\\
$\mathrm{HR@5 \uparrow}$ & $0.1422$ & $0.0625$ & $0.0027$ & $0.0029$ & $0.0123$ & $0.0823$ & $0.1304$ & $0.1294$ & $0.1107$ & $0.0867$ & $\underline{0.1532}$ & $0.1136$ & $0.1038$ & $0.1400$ & $\textbf{0.1667(+8.8\%)}$\\
$\mathrm{NDCG@5 \uparrow}$ & $0.1323$ & $0.0525$ & $0.0022$ & $0.0019$ & $0.0064$ & $0.0747$ & $0.1234$ & $0.1230$ & $0.1098$ & $0.0803$ & $\underline{0.1434}$ & $0.1002$ & $0.0970$ & $0.1290$ & $\textbf{0.1611(+12.3\%)}$\\
$\mathrm{repeat@10 \downarrow}$ & $-$ & $-$ & $0.89\%$ & $3.15\%$ & $\textbf{0.00\%}$ & $\textbf{0.00\%}$ & $9.02\%$ & $34.69\%$ & $\textbf{0.00\%}$ & $\textbf{0.00\%}$ & $\textbf{0.00\%}$ & $\textbf{0.00\%}$ & $\textbf{0.00\%}$ & $\textbf{0.00\%}$ & $1.00\%$\\
$\mathrm{OOD@10 \downarrow}$ & $-$ & $-$ & $61.21\%$ & $52.52\%$ & $11.91\%$ & $\textbf{0.00\%}$ & $8.13\%$ & $14.85\%$ & $\textbf{0.00\%}$ & $\textbf{0.00\%}$ & $\textbf{0.00\%}$ & $\textbf{0.00\%}$ & $\textbf{0.00\%}$ & $\textbf{0.00\%}$ & $\textbf{0.00\%}$ \\
\hline
\multicolumn{16}{c}{Dataset: Toys} \\
\hline

$\mathrm{HR@10 \uparrow}$ & $0.0589$ & $0.0389$ & $0.0031$ & $0.0039$ & $0.0354$ & $0.0405$ & $0.0473$ & $0.0438$ & $0.0498$ & $0.0371$ & $\textbf{0.0790}$ & $0.0515$ & $0.0596$ & $0.0657$ & $\underline{0.0714}$\\
$\mathrm{NDCG@10 \uparrow}$ & $0.0484$ & $0.0228$ & $0.0013$ & $0.0020$ & $0.0153$ & $0.0272$ & $0.0378$ & $0.0369$ & $0.0402$ & $0.0301$ & $\underline{0.0584}$ & $0.0394$ & $0.0437$ & $0.0508$ & $\textbf{0.0651(+11.5\%)}$\\
$\mathrm{HR@5 \uparrow}$ & $0.0529$ & $0.0276$ & $0.0021$ & $0.0019$ & $0.0191$ & $0.0311$ & $0.0426$ & $0.0407$ & $0.0418$ & $0.0327$ & $\underline{0.0652}$ & $0.0420$ & $0.0499$ & $0.0543$ & $\textbf{0.0686(+5.2\%)}$\\
$\mathrm{NDCG@5 \uparrow}$ & $0.0464$ & $0.0192$ & $0.0010$ & $0.0013$ & $0.0104$ & $0.0242$ & $0.0363$ & $0.0359$ & $0.0376$ & $0.0287$ & $\underline{0.0540}$& $0.0397$ & $0.0405$ & $0.0470$ & $\textbf{0.0643(+19.1\%)}$\\
$\mathrm{repeat@10 \downarrow}$ & $-$ & $-$ & $0.31\%$ & $2.10\%$ & $\textbf{0.00\%}$ & $\textbf{0.00\%}$ & $5.91\%$ & $29.50\%$ & $\textbf{0.00\%}$ & $\textbf{0.00\%}$ & $\textbf{0.00\%}$ & $\textbf{0.00\%}$ & $\textbf{0.00\%}$ & $\textbf{0.00\%}$ & $0.86\%$\\
$\mathrm{OOD@10 \downarrow}$ & $-$ & $-$ & $89.57\%$ & $90.99\%$ & $4.16\%$ & $\textbf{0.00\%}$ & $7.80\%$ & $37.00\%$ & $\textbf{0.00\%}$ & $\textbf{0.00\%}$ & $\textbf{0.00\%}$ & $\textbf{0.00\%}$ & $\textbf{0.00\%}$ & $\textbf{0.00\%}$ & $\textbf{0.00\%}$ \\
\bottomrule
\end{tabular}
}
\caption{Overall recommendation performance comparison on three datasets. Best results are in \textbf{bold}, second-best are \underline{underlined}; traditional recommenders serve as ID-based reference baselines.}
\label{tab:SeqRecResult_new}
\end{table*}

\subsubsection{Baselines}

We compare RecLM to 12 baselines grouped into four categories.
(1) Traditional ID-based sequential recommenders include SASRec~\citep{kang2018self} and GRU4Rec~\citep{hidasi2015session}, which operate purely on interaction sequences.
(2) Frozen LLM baselines include GPT-4o and Llama3-8B-Instruct, along with Llama3-cgen, a prompt-based constrained-generation variant of the latter.
(3) Finetuned LLMs comprise BIGRec~\citep{bao2023bi}, CtrlRec~\citep{DBLP:conf/acl/LuLZL0LX24}, and PALR~\citep{yang2023palr}.
Finally, (4) item-tokenizer LLMs include IDGenRec~\citep{10.1145/3626772.3657821}, SETRec~\citep{lin2025order}, LC-Rec~\citep{zheng2024adapting}, and LETTER~\citep{10.1145/3627673.3679569}, all built on Llama3-8B-Instruct.
Additional implementation details for all baselines are given in Appendix~\ref{append:baseline}.

\subsection{Overall Performance}

Table~\ref{tab:SeqRecResult_new} summarizes the overall comparison.
On the key reliability metric, OOD@10, all three RecLM variants strictly avoid out-of-domain items across all benchmarks, matching the best mapping- and tokenizer-based baselines.
This shows that, under a unified \texttt{<SOI>}-based grounding interface, both retrieval and constrained generation can enforce catalog fidelity rather than trading it off against recommendation quality.

In terms of accuracy, the constrained generation variants deliver the strongest performance overall: RecLM-cgen and RecLM-token consistently match or exceed the best non-ours LLM baselines.
RecLM-cgen maintains $\mathrm{repeat@10}=0$, and RecLM-token also keeps repetition low, indicating that the unified framework can simultaneously control OOD, repetition, and ranking quality.
RecLM-ret typically trails the constrained generation variants in ranking metrics but remains competitive with prior mapping-based methods.

Beyond quantitative metrics, a qualitative case study (Appendix~\ref{appendix:case_study}) shows that the Title Rewriter succinctly compresses verbose item metadata into clearer identifiers and that rewriting user-history titles can further improve recommendation quality.



\subsection{Ablation Study}
\label{subsection:ablation}

We choose {RecLM-cgen} for our ablation study as it incorporates the most newly proposed components. We analyze how \textsc{RecLM-cgen} components contribute to accuracy and control. We define six variants: \textbf{v0} is a finetuned Llama3-8B-Instruct that can emit \texttt{<SOI>} but uses unconstrained decoding; \textbf{v1} adds constrained generation; \textbf{v2} adds the scope-mask loss; \textbf{v3} additionally uses multi-round dialogue data; \textbf{v4} incorporates a TR trained with three reward components; and \textbf{full} uses the five-reward TR, representing our complete configuration.

\begin{table}[t]
\resizebox{\linewidth}{!}{
\begin{tabular}{ll|cccccc}
\toprule
Dataset & Metrics & v0 & v1 & v2 & v3 & v4 & full \\
\bottomrule
\multirow{6}{*}{Steam}
& $\mathrm{HR@10 \uparrow}$ & $0.0731$ & $0.0749$ & $0.0746$ & $0.0797$ & $\underline{0.0829}$ & $\textbf{0.0868}$ \\
& $\mathrm{NDCG@10 \uparrow}$ & $0.0396$ & $0.0406$ & $0.0410$ & $0.0433$ & $\underline{0.0447}$ & $\textbf{0.0456} $ \\
& $\mathrm{HR@5 \uparrow}$ & $0.0495$ & $0.0508$ & $0.0502$ & $0.0540$  & $\underline{0.0571} $ & $\textbf{0.0579} $ \\
& $\mathrm{NDCG@5 \uparrow}$ & $0.0320$ & $0.0329$ & $0.0332$ & $0.0360$  & $\textbf{0.0362} $ & $\underline{0.0361}$  \\
& $\mathrm{Repeat@10 \downarrow}$ & $2.33\%$ & $0.00\%$ & $0.00\%$ & $0.00\%$ & $0.00\%$ & $0.00\%$\\
& $\mathrm{OOD@10 \downarrow}$ & $1.75\%$ & $0.00\%$ & $0.00\%$ & $0.00\%$ & $0.00\%$ & $0.00\%$\\

\hline
\multirow{6}{*}{Movies}
& $\mathrm{HR@10 \uparrow}$ & $0.1331$ & $0.1400$ & $\underline{0.1443}$ & $0.1424$ & $0.1433 $ & $\textbf{0.1467}$ \\
& $\mathrm{NDCG@10 \uparrow}$ & $0.1240$ & $0.1269$ & $\underline{0.1318}$ & $0.1296$ & $\textbf{0.1329} $ & $0.1311 $ \\
& $\mathrm{HR@5 \uparrow}$ & $0.1297$ & $0.1334$ & $\underline{0.1396}$ & $0.1365$ & $\textbf{0.1400} $ & $\textbf{0.1400} $  \\
& $\mathrm{NDCG@5 \uparrow}$ & $0.1229$ & $0.1248$ & $\underline{0.1303}$ & $0.1277$ & $\textbf{0.1318} $ & $0.1290 $ \\
& $\mathrm{Repeat@10 \downarrow}$ & $39.26\%$ & $0.00\%$ & $0.00\%$ & $0.00\%$ & $0.00\%$ & $0.00\%$\\
& $\mathrm{OOD@10 \downarrow}$ & $17.48\%$ & $0.00\%$ & $0.00\%$ & $0.00\%$ & $0.00\%$ & $0.00\%$\\

\hline
\multirow{6}{*}{Toys}
& $\mathrm{HR@10 \uparrow}$ & $0.0400$ & $0.0581$ & $0.0605$ & $0.0642$ & $\textbf{0.0686} $ & $\underline{0.0657} $ \\
& $\mathrm{NDCG@10 \uparrow}$ & $0.0346$ & $0.0429$ & $0.0442$ & $0.0479$ & $\underline{0.0481} $ & $\textbf{0.0508}$  \\
& $\mathrm{HR@5 \uparrow}$ & $0.0380$ & $0.0475$ & $0.0496$ & $\underline{0.0534}$ & $\textbf{0.0543}$ & $\textbf{0.0543} $ \\
& $\mathrm{NDCG@5 \uparrow}$ & $0.0340$ & $0.0395$ & $0.0407$ & $\underline{0.0444}$  & $0.0435$ & $\textbf{0.0470}$ \\
& $\mathrm{Repeat@10 \downarrow}$ & $34.57\%$ & $0.00\%$ & $0.00\%$ & $0.00\%$ & $0.00\%$ & $0.00\%$ \\
& $\mathrm{OOD@10 \downarrow}$ & $35.85\%$ & $0.00\%$ & $0.00\%$ & $0.00\%$ & $0.00\%$ & $0.00\%$ \\
\toprule

\end{tabular}}
\caption{Ablation study on three datasets comparing the performance of six model variants.}
\label{tab:AblationStudy}
\end{table}

Table~\ref{tab:AblationStudy} shows that each component of RecLM-cgen yields incremental gains.
Introducing constrained generation (\textbf{v1}) consistently improves ranking metrics over \textbf{v0} while preserving low OOD and repetition rates.
The scope mask (\textbf{v2}) further boosts accuracy by matching training to prefix-tree decoding, and multi-round dialogue training (\textbf{v3}) improves robustness when recommendation is interleaved with general conversation.
Adding the TR module (\textbf{v4} and \textbf{full}) delivers the best or near-best accuracy, indicating that RL-optimized titles help the model use its constrained generation capacity more effectively.
Together, these trends illustrate how the unified RecLM design can be tuned along several axes—decoding constraints, training alignment, and title rewriting—to strengthen both reliability and recommendation quality.

\begin{table}[]
\resizebox{\linewidth}{!}{
\begin{tabular}{l|ccc|cc}
\toprule
& \multicolumn{3}{c|}{Response $R_1$} & \multicolumn{2}{c}{Response $R_2$} \\
\cmidrule{2-6}
Model & $\mathrm{HR@10 \uparrow}$ & $\mathrm{NDCG@10 \uparrow}$ & $\mathrm{CSN_{R_1}^{n=10} \uparrow}$ & $\mathrm{ACC_{gsm8k} \uparrow}$ & $\mathrm{CSN_{R_2}^{n=0} \uparrow}$\\
\bottomrule

\multicolumn{6}{c}{Dataset: Steam} \\
\hline
$\text{Llama3-cgen}$ & $0.0258$ & $0.0119$ & $0.717$ & $\textbf{0.676}$ & $\textbf{1.000}$ \\
$\mathrm{PALR}$ & $0.0629$ & $\underline{0.0364}$ & $-$ & $0.585$ & $-$ \\
$\mathrm{CtrlRec}$ & $\underline{0.0662}$ & $0.0349$ & $-$ & $0.022$ & $-$ \\
$\text{RecLM-ret}$ & $0.0508$ & $0.0257$ & $\underline{0.998}$ & $0.669$ & $0.987$ \\
$\text{RecLM-cgen}$ & $\textbf{0.0713}$ & $\textbf{0.0410}$ & $\textbf{1.000}$ & $\underline{0.673}$ & $0.990$ \\
$\text{RecLM-token}$ & $0.0480$ & $0.0221$ & $\textbf{1.000}$ & $0.660$ & $\underline{0.998}$ \\
\hline

\multicolumn{6}{c}{Dataset: Movies} \\
\hline
$\text{Llama3-cgen}$ & $0.0296$ & $0.0128$ & $0.703$ & $\underline{0.670}$ & $\textbf{1.000}$ \\
$\mathrm{PALR}$ & $0.1425$ & $0.1349$ & $-$ & $0.380$ & $-$ \\
$\mathrm{CtrlRec}$ & $0.1327$ & $0.1233$ & $-$ & $0.456$ & $-$ \\
$\text{RecLM-ret}$ & $0.1062$ & $0.0944$ & $\underline{0.998}$ & $0.653$ & $0.987$ \\
$\text{RecLM-cgen}$ & $\underline{0.1509}$ & $\underline{0.1388}$ & $\textbf{1.000}$ & $\textbf{0.703}$ & $\underline{0.988}$ \\
$\text{RecLM-token}$ & $\textbf{0.1524}$ & $\textbf{0.1441}$ & $\textbf{1.000}$ & $0.652$ & $\textbf{1.000}$ \\
\hline

\multicolumn{6}{c}{Dataset: Toys} \\
\hline
$\text{Llama3-cgen}$ & $0.0403$ & $0.0245$ & $0.396$ & $\underline{0.667}$ & $\textbf{1.000}$ \\
$\mathrm{PALR}$ & $0.0462$ & $0.0399$ & $-$ & $0.617$ & $-$ \\
$\mathrm{CtrlRec}$ & $0.0455$ & $0.0404$ & $-$ & $0.591$ & $-$ \\
$\text{RecLM-ret}$ & $0.0516$ & $0.0396$ & $\underline{0.998}$ & $0.640$ & $\textbf{1.000}$ \\
$\text{RecLM-cgen}$ & $\underline{0.0584}$ & $\underline{0.0484}$ & $\textbf{1.000}$ & $\textbf{0.718}$ & $\underline{0.998}$ \\
$\text{RecLM-token}$ & $\textbf{0.0644}$ & $\textbf{0.0545}$ & $\textbf{1.000}$ & $0.638$ & $\underline{0.998}$ \\
\toprule

\end{tabular}
}
\caption{Results of the control symbol study in the multi-turn dialogue setting.}
\label{tab:MultiRoundResult}
\end{table}

We further probe robustness under domain shift by training and evaluating RecLM-cgen across mismatched domains (e.g., training on one catalog and testing on another). Detailed cross-domain results are reported in the Appendix \ref{app:crossdomain}.


\subsection{Control Symbol Study} 
\label{sec:control_symbol_exp}

            

            

            



To assess the reliability of control symbol generation, we use a multi-turn dialogue setting to ensure that the final model can interact naturally in daily conversations and provide effective recommendations. We construct a three-turn dialogue that interleaves GSM8K-style math questions with a recommendation request. The first and third turns are math reasoning tasks, while the second asks for 10 item recommendations. We measure $CSN_{R_*}^{n=k}$, the proportion of responses $R_*$ that generate exactly $k$ \texttt{<SOI>} symbols (10 for the recommendation turn, 0 for non-recommendation turns).

As shown in Table~\ref{tab:MultiRoundResult}, both {RecLM-cgen} and {RecLM-token} achieve near-perfect $CSN$ scores, correctly generating 10 \texttt{<SOI>} symbols in recommendation turns and almost never emitting them in reasoning turns. Importantly, this high level of control is achieved without sacrificing recommendation accuracy, as evidenced by their competitive HR@10 and NDCG@10 scores across all datasets. This indicates the robustness of the control-symbol interface in our unified framework.

\subsection{General Tasks Evaluation}  

\begin{table}[]
\resizebox{\linewidth}{!}{
\begin{tabular}{ll|cccc}
\toprule
Dataset & Model & MMLU & GSM8K & CSQA & Humam-eval \\
\bottomrule
\rowcolor{gray!20} - & $\mathrm{Llama3}$ & $0.675$ & $0.781$ & $0.786$ & $0.640$ \\
\hline

\multirow{6}{*}{Steam}
& $\mathrm{BIGRec}$ & $0.632$ & $0.722$ & $0.737$ & $0.402$ \\
& $\mathrm{PALR}$ & $\textbf{0.659}$ & $\underline{0.745}$ & $\textbf{0.778}$ & $0.512$ \\
& $\mathrm{CtrlRec}$ & $0.646$ & $0.697$ & $0.764$ & $0.567$ \\
& $\text{RecLM-ret}$ & $0.653$ & $0.722$ & $0.762$ & $\underline{0.573}$ \\
& $\text{RecLM-cgen}$ & $\underline{0.657}$ & $\textbf{0.777}$ & $\underline{0.767}$ & $\textbf{0.591}$ \\
& $\text{RecLM-token}$ & $0.638$ & $0.676$ & $0.750$ & $0.530$ \\
\hline

\multirow{6}{*}{Movies}
& $\mathrm{BIGRec}$ & $0.609$ & $0.689$ & $0.711$ & $0.299$ \\
& $\mathrm{PALR}$ & $0.651$ & $\underline{0.747}$ & $0.747$ & $\underline{0.555}$ \\
& $\mathrm{CtrlRec}$ & $0.649$ & $0.729$ & $0.756$ & $0.549$ \\
& $\text{RecLM-ret}$ & $0.650$ & $0.501$ & $\underline{0.761}$ & $\underline{0.555}$ \\
& $\text{RecLM-cgen}$ & $\textbf{0.658}$ & $\textbf{0.772}$ & $0.756$ & $\textbf{0.579}$ \\
& $\text{RecLM-token}$ & $\underline{0.653}$ & $0.737$ & $\textbf{0.762}$ & $0.506$ \\
\hline

\multirow{6}{*}{Toys}
& $\mathrm{BIGRec}$ & $0.622$ & $0.661$ & $0.710$ & $0.445$ \\
& $\mathrm{PALR}$ & $\underline{0.645}$ & $\underline{0.728}$ & $0.737$ & $\underline{0.561}$ \\
& $\mathrm{CtrlRec}$ & $0.623$ & $0.721$ & $0.728$ & $\underline{0.561}$ \\
& $\text{RecLM-ret}$ & $\textbf{0.653}$ & $0.340$ & $\underline{0.754}$ & $\underline{0.561}$ \\
& $\text{RecLM-cgen}$ & $\textbf{0.653}$ & $\textbf{0.767}$ & $0.747$ & $\textbf{0.598}$ \\
& $\text{RecLM-token}$ & $0.640$ & $0.709$ & $\textbf{0.756}$ & $0.512$ \\
\bottomrule

\end{tabular}
}
\caption{General-task performance. Gray row: untuned Llama3-8B-Instruct.}
\label{tab:GeneralTaskResult}
\end{table}

Finally, we examine how aligning models to recommendation and grounding affects their broader abilities. We evaluate MMLU (5-shot), GSM8K (8-shot), CommonsenseQA (7-shot), and HumanEval (0-shot), covering comprehension, mathematics, commonsense reasoning, and code generation. As shown in Table~\ref{tab:GeneralTaskResult}, while there is a general slight decline compared to the untuned Llama3-8B-Instruct due to the alignment tax, our \texttt{RecLM} variants demonstrate strong resilience against catastrophic forgetting. RecLM-cgen consistently achieves superior performance among tuned models; it largely tracks Llama3's capabilities in mathematical reasoning and significantly outperforms other baselines in code generation, indicating that our grounding-oriented fine-tuning effectively preserves general reasoning and language understanding.

\section{Conclusion}

Advancing the application of LLMs to recommendation systems, this paper introduced \textsl{RecLM}, a unified framework that eliminates OOD recommendations through control tokens (\texttt{<SOI>}/\texttt{<EOI>}) and interchangeable modules: retrieval-based (\textsl{RecLM-ret}), constrained generation (\textsl{RecLM-cgen}), and item tokenization (\textsl{RecLM-token}). All variants achieve OOD@10 = 0. The unified view enables fair comparison of OOD-avoidance paradigms, driving scientific findings on their trade-offs. \textsl{RecLM} variants attain state-of-the-art recommendation accuracy and serves as a practical tool for real-world model training and deployment.

\section*{Limitations}

While {RecLM-cgen} and {RecLM-Token} demonstrate significant improvements in recommendation accuracy and successfully address the out-of-domain item generation problem, several limitations warrant further discussion and investigation.

\subsection{Inference Latency and Scalability}

The inference latency of LLM-based generative recommendations presents challenges for large-scale, real-time services that require millisecond-level response times. While our framework employs prefix-tree constrained decoding to reduce search space, the autoregressive nature of token-by-token generation remains computationally intensive. Future work should implement and benchmark specific optimization techniques: \textit{model distillation} to create smaller, specialized variants; \textit{speculative decoding} to accelerate generation; \textit{quantization-aware training} for efficient deployment; and \textit{hybrid architectures} that combine retrieval efficiency with generative refinement. These optimizations could make \textsc{RecLM} variants practical for industrial-scale applications.

\subsection{Evaluation Beyond Accuracy}

Our evaluation focused primarily on accuracy metrics (NDCG, Hit Rate), but comprehensive recommender system assessment requires examining diversity, fairness, and long-term user satisfaction. Future work should implement \textit{multi-objective optimization} during training, incorporating diversity constraints and fairness regularizers. For evaluation, we recommend adopting established metrics like \textit{intra-list diversity}, \textit{coverage}, and \textit{equity measures} across demographic segments. Additionally, \textit{longitudinal user studies} and \textit{online A/B testing} frameworks are needed to assess real-world impact beyond offline metrics. These enhancements would provide a more holistic assessment of {RecLM}'s practical utility.

\bibliography{custom}

\appendix

\section{Appendix}
\label{appendix:expertiment-details}
\subsection{Data Augmentation and Experiment Setup}

\label{appendix:data_aug}

During the training process of RecLM-ret, RecLM-cgen(Excluding the training of the TR) and RecLM-token, we employ an online data augmentation strategy. For each user’s historical interaction records, denoted as \(I_{history}^{(1...n)}\), we randomly sample a continuous segment \(I_{history}^{(a...b)}\) where \(1 \leq a < b \leq 10 < n\), to serve as the augmented training data. To construct the corresponding training labels for  \(I_{history}^{(a...b)}\), denoted as \(I_{rec}^{(1...k)}\), where \(k\) is a random integer between 1 and 10, we follow the method described in \citep{DBLP:conf/acl/LuLZL0LX24}. Specifically, \(I_{rec}^{(1)}\) corresponds to the next interaction item \(I_{history}^{(b+1)}\), while \(I_{rec}^{(2...k)}\) are provided by the teacher model SASRec based on \(I_{history}^{(a...b)}\).

Before the start of each epoch, data augmentation sampling is performed for every user. As a result, the training data for each epoch corresponds to the total number of users in the dataset, with online augmentation ensuring greater diversity in the training samples. During the testing phase, no data augmentation is applied. Instead, the number of test samples remains fixed at $10,000$.

\subsection{Baseline Details}
\label{append:baseline}
\paragraph{LLMs (frozen)} \begin{itemize} \item \textbf{GPT-4o}: The \textit{gpt-4o-2024-05-13} version accessed via Azure OpenAI. \item \textbf{Llama3}: The \textit{Llama3-8b-instruct} model, which also serves as the base model for our tuning. \item \textbf{Llama3-cgen}: A prompt-based variant where Llama3 is instructed to output a special symbol \texttt{<SOI>} before mentioning an item, triggering our constrained generation decoding. \end{itemize}

\paragraph{LLMs (finetuned)} \begin{itemize} \item \textbf{BIGRec}~\citep{bao2023bi}: Fine-tunes the LLM to generate item-related text, which is then mapped to the item corpus using an embedding model (BGE-M3). \item \textbf{CtrlRec}~\citep{DBLP:conf/acl/LuLZL0LX24}: Focuses on controllability using two-stage training (supervised fine-tuning and reinforcement learning). It employs SASRec as a teacher model for data augmentation during SFT. \item \textbf{PALR}~\citep{yang2023palr}: Relies on SFT to learn recommendation tasks. We explicitly enable the SASRec-based data augmentation for PALR in our experiments to match the performance gains demonstrated in~\citep{DBLP:conf/acl/LuLZL0LX24}. \end{itemize} CtrlRec and PALR both use SASRec-based data augmentation for fair comparison, which can be found in Appendix \ref{appendix:data_aug}.

\paragraph{LLMs (item tokenizer)} \begin{itemize} \item \textbf{IDGenRec}~\citep{10.1145/3626772.3657821}: Optimizes an item generator to dynamically adjust textual item IDs and utilizes a prefix tree for decoding. \item \textbf{SETRec}~\citep{lin2025order}: Introduces an order-agnostic set identifier paradigm. It integrates collaborative filtering and semantic information while employing a query-guided mechanism to enable simultaneous generation for enhanced efficiency. \item \textbf{LC-Rec}~\citep{zheng2024adapting}: Addresses the semantic gap by utilizing a vector quantization-based item indexing mechanism with uniform semantic mapping. It employs multi-faceted alignment tuning tasks to integrate collaborative and language semantics. \item \textbf{LETTER}~\citep{10.1145/3627673.3679569}: Learns a hierarchical item tokenizer through codebooks and decodes items using prefix trees. \end{itemize}


\begin{table}[t]
\centering
\resizebox{\linewidth}{!}{%
\begin{tabular}{lcccc} 
\toprule  
Dataset & \#Users & \#Items & \#Inters & \#Sparsity \\
\midrule  
Steam   & 10,000 & 11,726 & 170,000 & 99.85\% \\
Movies  & 10,000 & 34,452 & 170,000 & 99.95\% \\
Toys    & 10,000 & 49,985 & 170,000 & 99.96\% \\
\bottomrule 
\end{tabular}%
}
\caption{General statistics of the three datasets used in our experiments.}
\label{tab:DatasetStatistic}
\end{table}

\subsection{Additional Training Details}
\label{appendix:another_traing_details}
\paragraph{Backbone and Fine-tuning.}
Unless otherwise stated, all RecLM variants are initialized from Llama3-8B-Instruct.
User behavior sequences are truncated to at most 10 interactions and injected into instruction-style prompts as user profiles, with the maximum input and output lengths both set to 512 tokens.
We fine-tune all linear layers of the backbone using LoRA (PEFT) with the Adam optimizer, a learning rate of $1\times10^{-4}$, LoRA rank $r=16$, scaling factor $\alpha=8$, and a batch size of 2.
Training typically converges within 20 epochs.
The embeddings of the control symbols \texttt{<SOI>} and \texttt{<EOI>} are initialized as the average of the token embeddings for the phrases ``start of an item'' and ``end of an item'', respectively.
All quantitative results are averaged over five independent runs, and we report significance using paired tests with $p<0.05$.

\paragraph{Item Tokenization Setup.}
The item tokenizer is trained on all items that appear in the interaction logs.
For each item, we combine its collaborative embedding with its semantic embedding and feed them as inputs, while discrete item codes are learned in an unsupervised manner.
We use residual vector quantization with four codebooks, each containing 256 entries, so that every item is represented by a tuple of four codes.
The tokenizer is optimized by minimizing a reconstruction loss together with a commitment-style quantization loss to encourage stable code assignments; we set the weighting coefficient $\lambda_6 = 1.0$ and the commitment parameter $\beta = 0.25$ in all experiments.
To assess tokenization quality, we monitor both reconstruction error and collision rate, defined as the proportion of distinct items that are mapped to identical code tuples.

\paragraph{Reinforcement Learning Setup.}
Reinforcement learning data are constructed on-the-fly from raw user interaction sequences.
Given a sequence $I_{hist} = [i_1, i_2, \dots, i_N]$, we randomly sample a split point $t$ and use the prefix $I_{hist,<t}$ as the observed history, while the item $i_t$ is treated as the ground-truth target for reward computation and is never exposed to the model during generation.
The history is formatted as an instruction-style prompt asking the model to produce a top-$K$ recommendation list.
During RL training, we sample 16 candidate recommendation lists per prompt to enable relative reward normalization and stable optimization.
Unless otherwise specified, we use a sampling temperature of 1.0, a maximum generation length of 128 tokens, a learning rate of $1\times10^{-5}$, and a batch size of 32, with a KL regularization coefficient of 0.04.
For LoRA during RL, the rank and scaling factor are set to 16 and 32, respectively.
We fix the recommendation list length to $K=10$ and the maximum history length in RL to 20, matching the main-sequence setting described in the experiment section.

\subsection{Projection Layer of RecLM-ret}

In RecLM-ret, to align the hidden representation \(\mathbf{h}^{(i)}_{<SOI>}\) of the base model with the vector space of the pre-generated item embeddings \(\mathcal{E}\), we introduce a projection layer. Its formulation is shown in Equation~\ref{eq:project_layer}.

\begin{equation}\label{eq:project_layer}
\text{proj}_{\phi}(\mathbf{h}^{(j)}_{<SOI>})=GELU(\mathbf{h}^{(j)}_{<SOI>}\cdot \mathbf{W}_1)\cdot \mathbf{W}_2
\end{equation}

Here, $\mathbf{W}_1 \in \mathbb{R}^{d \times \frac{d}{2}}$ and $\mathbf{W}_2 \in \mathbb{R}^{\frac{d}{2} \times c}$ constitute the trainable parameters $\phi$ of the project layer. $d$ is the dimension of base model. $c$ is the dimension of item embeddings \(\mathcal{E}\).

\subsection{Prompt Settings in RecLM-ret and RecLM-cgen}
\label{app:Prompts}
We provide the prompts in Listing~\ref{lst:training_prompts} which are used to convert user behaviors <$I^{(1...n)}_{history}$, $I^{(1...k)}_{rec}$> into Supervised Fine-Tuning data samples <Instruction:$X$, Response:$Y$>. Io increase the data diversity, we use four prompt templates.

\subsection{Prefix Tree Structure and Constrained Generation}
We construct a prefix tree based on the item titles within the given recommendation domain. This prefix tree is represented as $Node={n_1,...,n_a}$ and $Children={C_1,...,C_a}$, where $a$ is the number of nodes in the prefix tree, and $C_i$ is a set indicating the child nodes of node $n_i$. To avoid recommending the same item multiple times within the same response, we record the number of leaf nodes under the subtree of each node in this prefix tree as $L={l_1,...,l_a}$ (where $l_i$ is the number of leaf nodes in the subtree corresponding to node $n_i$, indicating the maximum number of times node $n_i$ can be accessed within a single response).

At a certain generation step during the inference phase, the input token sequence is $X=[t_1,...,t_i]$, and the token sequence of the generated response is $Y=[t_{i+1},...t_{i+j}]$. We look for the most recent control token in sequence $Y$. If the most recent control token is \texttt{<SOI>} (Start of Item), then we activate the constrained decoding strategy. If the most recent control token is \texttt{<EOI>} (End of Item) or no control token has been generated yet, the constrained decoding strategy is not activated.

When the constrained decoding strategy is activated (the most recent control token $t_{i+k}$ is \texttt{<SOI>}, where $1 \leq k \leq j$), we first need to count the access times of the recommended items in the generated response at their corresponding nodes in the prefix tree, denoted as $V={v_1,...,v_a}$, where $v_i \leq l_i$. Next, we locate the corresponding node $n_b$ in the prefix tree based on the sequence $[t_{i+k},...,t_{i+j}]$ and obtain the set of candidate next tokens $C_b$. To avoid generating duplicate items, we exclude nodes in $C_b$ whose access times have reached the maximum access times, resulting in $C_{b}^{'}$ as the final candidate set for token $t_{i+j+1}$. We set the logit values of tokens outside $C_{b}^{'}$ to negative infinity to prevent them from being generated.

\begin{figure}[ht]
    \centering
    \includegraphics[width=0.5\textwidth]{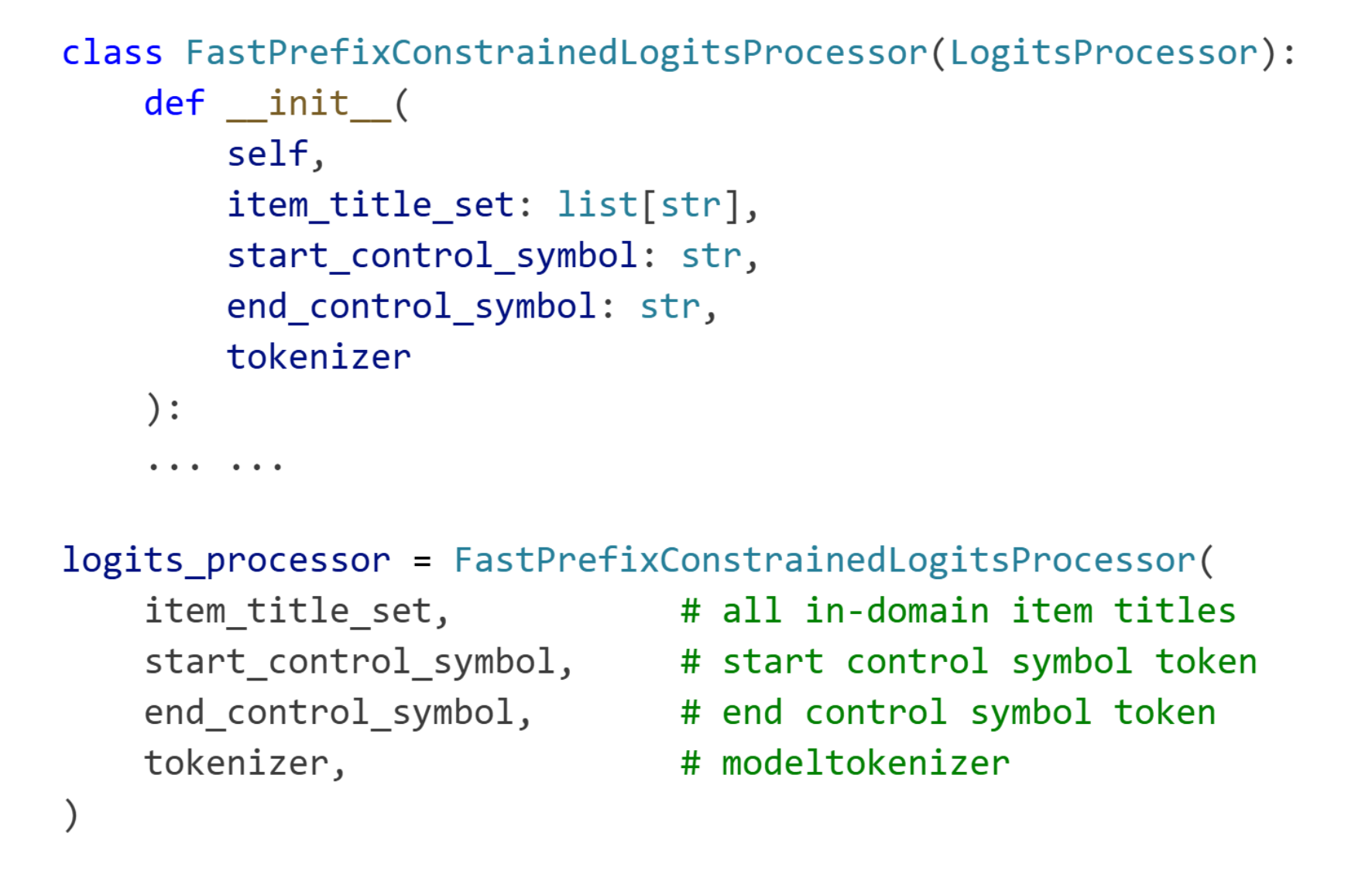}
    \caption{Example implementation of RecLM-cgen during inference. Only minimal code modifications are required to integrate the constrained generation mechanism}
    \label{fig:torch_code}
\end{figure}

A key feature of RecLM-cgen is its simplicity in inference, as demonstrated in Figure~\ref{fig:torch_code}. 

\subsection{Discussions on RecLM-cgen vs. RecLM-ret}
\label{sec:discussion}
In this section, we provide some theoretical perspective on why RecLM-cgen tends to achieve higher recommendation accuracy than RecLM-ret.

The main difference in the paradigms of RecLM-cgen and RecLM-ret is \textbf{Single-Stage Generation} vs. \textbf{Two-Stage Retrieval}. RecLM-ret relies on a two-step process:

\begin{enumerate}
\item Generate a special \texttt{<SOI>} token.
\item Perform a similarity-based lookup in an external embedding index to select the item.
\end{enumerate}

This split can degrade accuracy in two ways. First, any mismatch between the model’s hidden-state embedding and the item corpus embeddings may select a suboptimal item. Second, because the retrieval is effectively an external “hard choice,” it does not benefit from token-by-token language modeling feedback, i.e., once \texttt{<SOI>} is emitted, the model’s subsequent text has no bearing on which item is retrieved.

RecLM-cgen, by contrast, never leaves its native autoregressive process. Once \texttt{<SOI>} is produced, the model continues to generate tokens for the item title, except it restricts that token distribution to valid item titles stored in a prefix tree. In other words, each token that forms the recommended item is chosen within the model’s next-token probabilities. On the one hand, there is no embedding mismatch. The model’s hidden state directly translates into item-token predictions, rather than relying on an external embedding query.  On the other hand, it is using a unified generative signal. Every generated token refines the item selection process. The model’s full contextual understanding, such as user preferences, conversation history, etc., affects which item tokens appear.  

Mathematically, we can view RecLM-ret as factorizing the recommendation process into:
\begin{equation}  
P(\text{item}) \,\approx\, \mathrm{NN}\bigl(\phi(\mathbf{h}_{\texttt{<SOI>}}), \mathbf{E}\bigr)   
\end{equation}  
where $\phi$ is a projection of the model’s hidden state, and $\mathbf{E}$ is the precomputed item embedding base. Small errors in $\phi(\mathbf{h}_{\text{SOI}})$ can lead to suboptimal recommendations.

Conversely, RecLM-cgen effectively implements:
\begin{equation}  
P(\text{item} \mid \text{context}) \,=\,\prod_{i} P_{\theta}(w_i \mid w_{<i}, \text{context}),  
\end{equation}  
with the prefix-tree constraint filtering out invalid tokens. This direct language modeling over the item strings harnesses the entire generative capacity of the LLM, typically converging to higher recommendation accuracy during training. On the one hand, the model is trained to maximize the probability of each correct token in an item title, directly linking language modeling loss to better item predictions; on the other hand, there is no discontinuity between item selection and item-text generation, each token reflects the same internal distribution that learned the user’s context.

\section{Title Rewriter Training Details}
\label{sec:TR_info}

\subsection{The Design of the Reward}
\label{reward_design}
Since title rewriting lacks a single optimal answer, we train the TR using the GRPO algorithm with reinforcement learning. We design five reward functions to guide generation quality, focusing on recommendation effectiveness, LLM compatibility, appropriate length, and semantic distinctiveness from yet closeness to the original.

\subsubsection{Recommendation-Oriented Rewards}

The first category of rewards is designed to enhance the performance of the recommendation system. We consider two perspectives in this context: item-to-item similarity and user-to-item alignment. Since the generated titles serve as item identifiers, it is important that they should help in efficient item discovery and user profiling.

\paragraph{Item-to-Item Reward}

For the item-to-item perspective, the generated short titles should improve the identification of similar items. This is motivated by the observation that recommendation systems typically recommend items based on similarity. To quantify item similarity, we construct a contribution matrix $C \in \mathbb{R}^{N \times N}$, where each entry $C_{ij}$ represents the number of users who have interacted with both item $i$ and item $j$, and $N$ is the number of items. To mitigate the influence of popularity bias, we normalize the similarity scores as follows:

\begin{equation}
S_{ij} = \frac{C_{ij}}{|C_{i\cdot}| \cdot |C_{\cdot j}|}
\end{equation}

\noindent Here, $|C_{i\cdot}|$ and $|C_{\cdot j}|$ denote the number of users who interacted with items $i$ and $j$, respectively.

During training, the TR samples multiple candidate titles for each item. These rewritten titles are combined with the item descriptions and embedded using the BGE-M3 model. We then compute the cosine similarity between the resulting embedding and the embeddings of other original items, and compare the similarity-based ranking to the normalized similarity scores derived from the contribution matrix. Specifically, we select the top-10 items most similar to the current item based on the original contribution scores. After rewriting, we assess how the similarity-based ranking of these items changes by computing the Spearman’s rank correlation coefficient between the original and updated rankings. 

\paragraph{User-to-Item Reward}
\label{sec:u2ir}
From the perspective of user-item interaction, the objective is to evaluate whether the generated item titles better reflect user preferences. Each user's interaction history is represented by $K$ items. During the training of the TR, the TR is required to generate new titles for all $K$ items in a single sampling step. Based on these rewritten titles, we use RecLM-ret to obtain the user's final embedding by computing the user representation from the updated item history via a projection layer. We then assess the alignment between this new user embedding and the embedding of a target item (i.e., the item to be recommended). Specifically, we calculate cosine similarities between the user embedding and all item embeddings, determine the ranking position of the target item among these, and convert this ranking into an appropriate reward value.

\subsubsection{Decoding Complexity}

To ensure that the generated titles are easy to interpret and process by language models, we assess their decoding complexity using Perplexity (PPL). A lower PPL indicates that the title is more natural and easier for a language model to decode. During training, we obtain each user's interaction history and require the TR to rewrite the target item's title. The user’s interaction history and the rewritten target item are then concatenated into a single input sequence. We compute the log-likelihood of the target item portion within this sequence and use it to calculate the PPL of the newly generated item title.

\subsubsection{Conciseness Reward}

In addition to semantic quality, conciseness is another desirable property of generated titles. We define a length-based reward to encourage the generation of more concise titles. This reward is computed by taking the ratio of the length of the newly generated title to that of the original title, and then converting this ratio into a reward value, where lower ratios correspond to higher rewards. This design aims to minimize the length of the rewritten title, thereby promoting conciseness.

\subsubsection{Discriminative Power Reward}

Finally, the generated title should be distinguishable from the titles of semantically similar items. To evaluate this property, we design a discrimination task where the generated title is used as a prompt to a language model, which is then asked to identify the correct original title from a set of four candidates: the true original title and the titles of the three most similar items. These similar titles are selected based on cosine similarity between item embeddings. A higher identification accuracy indicates that the generated title is more distinctive and recognizable. The prompt template at this stage is as follows:










\subsection{Parameter Settings for Training}

The TR is trained using the GRPO with five reward functions. The base model is Llama3-8B-Instruct, fine-tuned with LoRA (rank 16, $\alpha = 32$). vLLM is used for efficient generation, with 4 completions sampled per input and a temperature of 0.6. Training is conducted with bfloat16 precision, a learning rate of $5 \times 10^{-6}$, for 1 epoch, using a per-device batch size of 4 and gradient accumulation steps of 2. Optimizer and model states are offloaded to reduce memory usage. Training was performed on 2 NVIDIA H100 GPUs and took approximately 6 hours to complete.

The optimal weights for the reward functions were determined empirically for each dataset. consistently, we set the weights for Conciseness (CR), Discriminative Power (DPR), and Item-to-Item Similarity (I2I) to 1 across all datasets. The weights for Decoding Complexity (DC) and User-to-Item Alignment (U2I) were tuned specifically: for the Steam dataset, we utilized $w_{DC}=2.5$ and $w_{U2I}=3$. For both the Movies and Toys datasets, we adopted $w_{DC}=2$ and increased the alignment weight to $w_{U2I}=4$.

During training, four reward functions—Item-to-Item Similarity, Decoding Complexity, Conciseness, and Discriminative Power—are applied to the STR data, while the User-to-Item Alignment reward is excluded. In contrast, the GTR data training exclusively employs the User-to-Item Alignment reward, as it is more consistent with the group-level modeling objective.

\subsection{Model Selection during Training}
\label{model_select}
The TR module is built upon the Llama3-8B-Instruct model as its base architecture.
For semantic embedding of rewritten titles and item descriptions, we employ the BGE-M3 model.
To ensure consistency across different reward components, the decoding complexity and discriminative power evaluation stages also rely on Llama3-8B-Instruct.

\subsection{Prompt Settings in TR}

During the GRPO training phase of TR, we design two types of tasks based on the requirements of the title rewriting objective. The first task requires the TR to rewrite a single title, referred to as single-title rewriting(STR).  The second task involves rewriting a group of titles in a single pass and producing them in order, referred to as group-title rewriting(GTR). The prompt settings for both tasks are shown in the Figure~\ref{fig:prompt_templates}.

\begin{figure}[t]
\centering
\small 

\begin{promptbox}{Prompt Template: Single Title Rewriting (STR)}
\textbf{Item's Information:}
\begin{itemize}[leftmargin=*, nosep, label={}] 
    \item Title: \texttt{\{item\_title\}}
    \item Description: \texttt{\{item\_description\}}
    \item Category: \texttt{\{item\_tags\}}
\end{itemize}

\vspace{0.5em}
\textbf{Instructions:}
\begin{enumerate}[leftmargin=*, nosep] 
    \item Rewrite the title to be concise, clear, and descriptive.
    \item Ideally shorter than the original, capturing key features.
    \item Ensure the new title flows naturally.
    \item \textbf{Output ONLY the rewritten title. No explanations.}
\end{enumerate}
\end{promptbox}

\vspace{0.5em} 

\begin{promptbox}{Prompt Template: Group Title Rewriting (GTR)}
\textbf{Input List:} \texttt{\{item\_list\}}

\vspace{0.5em}
\textbf{Instructions:}
\begin{itemize}[leftmargin=*, nosep]
    \item Rewrite each title in the list to be concise and descriptive.
    \item Keep it short, capturing key features.
    \item \textbf{Output strictly in the following format:}
    \begin{itemize}[nosep]
        \item 1. [new title]
        \item 2. [new title]
        \item \dots
    \end{itemize}
    \item \textbf{Do not include extra text.}
\end{itemize}
\end{promptbox}

\caption{Prompt templates used for the STR and GTR tasks. Variable placeholders are denoted in brackets.}
\label{fig:prompt_templates}
\end{figure}



\section{Additional Experiments}

\subsection{Prefix Tree Construction and Complexity}
\label{app:prefix_tree}
The prefix tree used in RecLM-cgen (and RecLM-token) is built once offline using a trie over all (rewritten) item titles. 
The construction cost scales approximately linearly with the total number of title tokens, and during inference the decoding cost depends on the tree depth (i.e., title length) rather than the catalog size, since only the children of the current trie node are considered when masking logits. 
As summarized in Table~\ref{tab:prefix_tree_scalability}, scaling the catalog from 10k to 1M items leads to only a marginal increase in end-to-end inference time, confirming that constrained generation remains efficient even for very large item corpora.

\begin{table}[t]
    \centering
    \resizebox{\columnwidth}{!}{
    \begin{tabular}{lcc}
        \toprule
        Catalog & Build (s) & Inference (10k users) \\
        \midrule
        10k   & 0.08   & 8 m 37 s \\
        100k  & 2.71   & 8 m 44 s \\
        1M    & 29.78  & 8 m 45 s \\
        \bottomrule
    \end{tabular}
    }
    \caption{Scalability of prefix tree construction and constrained generation on catalogs of different sizes.}
    \label{tab:prefix_tree_scalability}
\end{table}

\subsection{Cold-start Study}
In this section, we conducted experiments to study the cold start capability of the model. For each dataset, we include 10k new users who have not been included in the training set, and limit their history randomly within the length range 4 to 10. The target items for all these 10k users are not included in the training set, which simulate the cold-start item scenario. The results are shown in Table~\ref{tab:ColdStartStudy}.


\begin{table}[t]
\centering

\resizebox{\linewidth}{!}{%
\begin{tabular}{llcc} 
\toprule
Dataset & Model & HR@10 & NDCG@10 \\
\midrule
\multirow{2}{*}{Steam} & Llama3-8b-cgen & 0.0039 & 0.0021 \\
 & RecLM-cgen & 0.0086 & 0.0057 \\
\midrule
\multirow{2}{*}{Movies} & Llama3-8b-cgen & 0.0162 & 0.0089 \\
 & RecLM-cgen & 0.0422 & 0.0291 \\
\midrule
\multirow{2}{*}{Toys} & Llama3-8b-cgen & 0.0202 & 0.0100 \\
 & RecLM-cgen & 0.0313 & 0.0183 \\
\bottomrule
\end{tabular}%
}
\caption{Cold-start performance comparison on the three datasets.}
\label{tab:ColdStartStudy}
\end{table}

\subsection{Cross-domain Recommendation}
\label{app:crossdomain}

\begin{table}[]
\resizebox{\linewidth}{!}{
\begin{tabular}{lll|cc}
\toprule
$D_{train}$ & $D_{test}$ & Model & $\mathrm{HR@10 \uparrow}$ & $\mathrm{NDCG@10 \uparrow}$ \\
\bottomrule

\multirow{5}{*}{Toys}
& \multirow{5}{*}{Movies} & {Llama3-cgen} & $0.0246$ & $0.0106$ \\
\cline{3-5}
& & $\mathrm{Our_{cg}}$ & $0.0743$ & $0.0651$ \\
& & $\mathrm{Our_{cg+sm}}$ & $\underline{0.0953}(+28.26\%)$ & $\underline{0.0817} (+25.50\%)$ \\
\cline{3-5}
& & $\mathrm{Our_{cg+mr}}$ & $0.0745$ & $0.0648$ \\
& & $\mathrm{Our_{cg+sm+mr}}$ & $\textbf{0.1170 (+57.05\%)}$ & $\textbf{0.1029 (+58.80\%)}$ \\
\hline

\multirow{5}{*}{Movies}
& \multirow{5}{*}{Toys} & {Llama3-cgen} & $0.0354$ & $0.0153$ \\
\cline{3-5}
& & $\mathrm{Our_{cg}}$ & $0.0503$ & $0.0384$ \\
& & $\mathrm{Our_{cg+sm}}$ & $\textbf{0.0572 (+13.72\%)}$ & $\textbf{0.0447 (+16.41\%)}$ \\
\cline{3-5}
& & $\mathrm{Our_{cg+mr}}$ & $0.0481$ & $0.0366$ \\
& & $\mathrm{Our_{cg+sm+mr}}$ & $\underline{0.0527} (+9.56\%)$ & $\underline{0.0414} (+13.11\%)$ \\
\toprule
\end{tabular}
}
\caption{Cross-domain recommendation results. Our proposed models progressively integrate semantic memory (sm) and multi-round reasoning (mr) on top of base conversational generation (cg). }
\label{tab:ZeroShotResult}
\end{table}

We additionally evaluate the cross-domain behavior of RecLM-cgen.
Table~\ref{tab:ZeroShotResult} reports two zero-shot settings: training on Toys and testing on Movies, and training on Movies and testing on Toys.
For clarity, we group RecLM-cgen variants using subscripts, where $Our$ denotes the base configuration (v0 in  section \ref{subsection:ablation}), and subscripts $cg$, $sm$, and $mr$ indicate the inclusion of constrained generation, scope mask, and multi-round conversation training, respectively.
Across both cross-domain scenarios, incorporating the scope mask component ($sm$) consistently yields better performance than its counterparts without $sm$, indicating that aligning the training objective with the constrained decoding space improves robustness under domain shift.

\subsection{Inference Speed on RecLM-cgen}
\begin{table}[ht]
    \centering
    \resizebox{\linewidth}{!}{
        \begin{tabular}{ll|ccccc}
        \toprule
        \multirow{2}{*}{Dataset} & \multirow{2}{*}{cg} & \multirow{2}{*}{$\text{Token}_{in}$} & \multirow{2}{*}{$\text{Token}_{out}$} & $\text{Speed}_{.avg}$ & $\text{Search Time}_{in}$ & $\text{Search Time}_{out}$ \\
         & & & & {(token/s)} & {(ms/token)} & {(ms/token)} \\
        \bottomrule
        \multirow{2}{*}{Steam} & w/ & 7726 & 7552 & 35.0385 & 1.0725 & 0.3234 \\
         & w/o & 7872 & 7552 & 36.6996 & - & - \\
        \hline
        \multirow{2}{*}{Movies} & w/ & 12970 & 7552 & 34.5347 & 1.4535 & 0.3221 \\
         & w/o & 11900 & 7552 & 36.3846 & - & - \\
        \hline
        \multirow{2}{*}{Toys} & w/ & 20838 & 7552 & 34.0883 & 1.9922 & 0.3237 \\
         & w/o & 19910 & 7552 & 36.9466 & - & - \\
        \toprule
        \end{tabular}
    }
    \caption{Computation cost analysis of constrained generation (cg) during inference across three datasets. Results are compared between models with (w/) and without (w/o) constrained generation. }
    \label{tab:SearchTimeCost}
\end{table}

To illustrate that the constrained generation does not cause significant latency on the LLM inference, we conduct an inference throughput experiment. We select 128 test samples from the test set of three datasets, generating 10 item recommendations per test sample. The model is deployed using the Hugging Face Transformers library\footnote{\url{https://github.com/huggingface/transformers}} on a single A100 GPU (40GB), with an inference batch size set to $1$. We used 5 test samples for warm-up and ignored the time it took to generate the first token. We then aggregate the number of inner prefix tree tokens ($Token_{in}$) and outer prefix tree tokens ($Token_{out}$), calculating the average search time for both token types in the settings. Here search time corresponds to the operation to determine the valid space in next token decoding. Table~\ref{tab:SearchTimeCost} shows the average results of 5 repeated experiments, numbers are aggregated from the response text of the 128 test samples, we report both settings with and without constrained generation. 

For the Steam dataset, with constrained generation enabled, a total of 7,726 inner tokens and 7,552 outer tokens are generated. The average generation speed is 35.0385 tokens/second. The average search time for inner tokens is 1.0725 ms/token, while for outer tokens, it is 0.3234 ms/token. As the length of item titles increases from the Steam dataset (6.0359 tokens/item) to the Movies dataset (10.1328 tokens/item) and further to the Toys dataset (16.2797 tokens/item), the search time for outer tokens remains stable, whereas the search time for inner tokens gradually increases.

\subsection{In Context Learning Study}
We conducted experiments comparing the performance of contextual learning with RecLM-cgen. In Table~\ref{tab:ICLStudy}, Llama3-8b-cgen refers to the unfine-tuned LLM with constrained generation enabled. We observed that performance actually drops when using a 5-shot setting. This finding aligns with existing literature\citep{yao2023knowledge}. The reason is that LLMs can become severely biased towards the provided few-shot examples. Unless these examples are dynamically retrieved by a recommender model tailored to the current data sample, improvements from few-shot strategies are unlikely. Therefore, we recommend fine-tuning the LLM to some extent for better recommendation performance.


\begin{table}[t]
\centering

\resizebox{\linewidth}{!}{%
\begin{tabular}{llcc} 
\toprule
\textbf{Dataset} & \textbf{Model} & \textbf{HR@10} & \textbf{NDCG@10} \\
\midrule
\multirow{3}{*}{Steam} & Llama3-8b-cgen (0-shot) & 0.0261 & 0.0125 \\
 & Llama3-8b-cgen (5-shot) & 0.0211 & 0.0112 \\
 & \textbf{RecLM-cgen (Ours)} & \textbf{0.0797} & \textbf{0.0433} \\
\midrule 
\multirow{3}{*}{Movies} & Llama3-8b-cgen (0-shot) & 0.0246 & 0.0106 \\
 & Llama3-8b-cgen (5-shot) & 0.0121 & 0.0059 \\
 & \textbf{RecLM-cgen (Ours)} & \textbf{0.1424} & \textbf{0.1296} \\
\midrule
\multirow{3}{*}{Toys} & Llama3-8b-cgen (0-shot) & 0.0354 & 0.0153 \\
 & Llama3-8b-cgen (5-shot) & 0.0303 & 0.0137 \\
 & \textbf{RecLM-cgen (Ours)} & \textbf{0.0642} & \textbf{0.0479} \\
\bottomrule
\end{tabular}%
}
\caption{In-context learning (ICL) comparison across three datasets.}
\label{tab:ICLStudy}
\end{table}

\subsection{Sensitivity Analysis of Reward Weights}

This section provides a sensitivity analysis of the reward weight configuration used in the TR module.
We vary the five reward weights in Eq.~(4) while keeping all other training settings fixed, and report performance on the Steam dataset.

Table~\ref{tab:sensitivity_color} shows that the model performance remains relatively stable across different weight combinations.
Although certain configurations yield marginally higher values on specific metrics, no single weight dominates the performance, and reasonable variations do not lead to significant degradation.
This indicates that the TR module does not rely on finely tuned hyperparameters and is robust to moderate changes in reward weighting.

Based on this analysis, we use a balanced default configuration for all reported experiments.

\newcolumntype{a}{>{\columncolor{gray!20}}c}

\begin{table}[t]
\centering
\small
\setlength{\tabcolsep}{2.8pt}

\begin{tabular}{ccccc|cccc} 
\toprule
\multicolumn{5}{c|}{\textbf{Component Weights}} & \multicolumn{4}{c}{\textbf{Evaluation Metrics}} \\ 
\cmidrule(r){1-5} \cmidrule(l){6-9}
\rowcolor{gray!10}
DC & CR & DPR & I2I & U2I & H@10 & N@10 & H@5 & N@5 \\ 
\midrule
1 & 1 & 1 & 1 & 1 & 0.0820 & 0.0412 & 0.0560 & 0.0328 \\
1 & 1 & 1 & 1 & 2 & 0.0900 & 0.0490 & 0.0630 & 0.0404 \\
1 & 1 & 1 & 1 & 3 & 0.0780 & 0.0390 & 0.0530 & 0.0310 \\
2 & 1 & 1 & 1 & 3 & \textbf{0.0930} & 0.0485 & \textbf{0.0640} & 0.0384 \\
3 & 1 & 1 & 1 & 3 & 0.0890 & 0.0472 & 0.0580 & 0.0373 \\
1 & 1 & 2 & 2 & 3 & 0.0920 & \textbf{0.0496} & \textbf{0.0640} & \textbf{0.0406} \\
1 & 2 & 1 & 1 & 3 & 0.0860 & 0.0418 & 0.0500 & 0.0317 \\
1 & 3 & 1 & 1 & 3 & 0.0820 & 0.0436 & 0.0570 & 0.0354 \\
\bottomrule
\end{tabular}
\caption{Sensitivity analysis showing the impact of different weight configurations. H@K denotes HR@K and N@K denotes NDCG@K.}
\label{tab:sensitivity_color}
\end{table}

\section{Case Study}

\label{appendix:case_study}

\subsection{Title Rewriting for Enhanced Recommendation}



\begin{table}[t]
\centering
\small 
\newcolumntype{L}{>{\raggedright\arraybackslash}X}

\begin{tabularx}{\columnwidth}{l L}
\toprule
\textbf{Field} & \textbf{Content} \\
\midrule
\multicolumn{2}{l}{\textit{\textbf{Case 1: Steam Games}}} \\
\midrule
Original Title & Ukrainian Ball in Search of Gas \\
Description & ... our hero touched it with a gorilka and turned into a ball! ... You play as a man turned into a ball and \textbf{you need to steal all the gas from the forest at all costs} ... \\
\textbf{Rewritten} & \textbf{Gas Quest} \\

\midrule
\multicolumn{2}{l}{\textit{\textbf{Case 2: Movies}}} \\
\midrule
Original Title & 55 Days at Peking VHS \\
Description & Diplomats, soldiers and other representatives ... fend off the siege of the International Compound in Peking \textbf{during the 1900 Boxer Rebellion} ... \\
\textbf{Rewritten} & \textbf{55 Days at Peking (1900 Boxer Rebellion)} \\

\midrule
\multicolumn{2}{l}{\textit{\textbf{Case 3: Toys}}} \\
\midrule
Original Title & SwimWays Toypedo Revolution - Colors May Vary \\
Description & \textbf{The SwimWays Toypedo Revolution dive toy} rockets through the pool up to 30 feet with amazing hydrodynamic action! ... \\
\textbf{Rewritten} & \textbf{SwimWays Toypedo Dive Toy} \\

\bottomrule
\end{tabularx}
\caption{Case study of the TR module.}
\label{tab:case_study_refined}
\end{table}

As shown in Table~\ref{tab:case_study_refined}, our case study illustrates that the TR module effectively combines item titles and descriptions to retain key information while removing redundancy. For example, it rewrites the complex game title "Ukrainian Ball in Search of Gas" as the concise "Gas Quest," adds contextual details to movies like "55 Days at Peking (1900 Boxer Rebellion)" for clearer disambiguation, and simplifies toy product names by dropping irrelevant phrases like "Colors May Vary" while preserving brand identity. These enhancements result in more concise, readable, and informative identifiers, thereby improving LLM comprehension.

\subsection{Additional Case Study on Cross-domain Generalization}

Table~\ref{tab:case_study_cross_domain} presents an additional qualitative case study to further examine the cross-domain and cold-start behavior of the proposed framework.
Unlike standard recommendation settings, cross-domain scenarios involve items that are entirely absent from the training interactions, making historical collaborative signals unavailable.

In this example, a RecLM-cgen model trained exclusively on the Steam dataset is prompted with an item from the Movies domain.
Despite the domain shift and the lack of domain-specific interaction data, the model generates a coherent and semantically accurate description of the unseen item.
This observation suggests that the model relies primarily on semantic cues derived from item titles rather than memorizing domain-specific interaction patterns.



\begin{table}[t]
\centering
\small 
\newcolumntype{L}{>{\raggedright\arraybackslash}X}

\begin{tabularx}{\linewidth}{@{}l L@{}} 
\toprule
\textbf{Field} & \textbf{Content} \\
\midrule
Item Title & \textit{Take the Money and Run} VHS \\
\addlinespace[0.5em] 

Original Desc. & Woody Allen's feature-film debut, \emph{Take the Money and Run}, is a mockumentary combining sight gags, sketch-like scenes, and stand-up humor. Allen plays Virgil Starkwell, a music-loving nebbish who turns to a life of crime at an early age \dots \\
\addlinespace[0.5em]

Generated Desc. & \emph{Take the Money and Run} is a 1969 American comedy film directed by Woody Allen. The film stars Allen and Janet Margolin and is known for its fast-paced humor and mockumentary style. It is considered one of Allen's early works and showcases his distinctive comedic voice. The film is available in VHS format through various retailers. \\
\bottomrule
\end{tabularx}
\caption{Cross-domain qualitative case study demonstrating the model's generation capabilities.}
\label{tab:case_study_cross_domain}
\end{table}

\subsection{Impact of Title Rewriting on Recommendation Outcomes}

\begin{table*}[p]
\centering

\renewcommand{\arraystretch}{0.85} 
\scriptsize 

\resizebox{\textwidth}{!}{%
\begin{tabular}{|c|c|p{5.5cm}|p{2.5cm}|p{5.5cm}|c|}
\hline
\textbf{Case} \rule[-1.5ex]{0pt}{4ex} & \textbf{Stage} & \textbf{User History} & \textbf{Target Item} & \textbf{Top-10 Recommendations} & \textbf{Hit?} \\
\hline

\multirow{22}{*}{\rotatebox{90}{\textbf{Case 1}}} 
& \multirow{11}{*}{\rotatebox{90}{Before}} 
& \begin{itemize}[leftmargin=*, nosep, itemsep=-1pt, label=\textbullet] 
    \item The Hobbit: The Battle of the Five Armies [DVD] [2015]
    \item Tinker Bell and the Legend of the Neverbeast
    \item 55 Days at Peking VHS
    \item The Incredibles (Mandarin Chinese Edition)
    \item The Legend of Longwood
    \item Batman vs. Robin
    \item The Last Witch Hunter Digital
    \item Batman: Bad Blood
    \item Justice League vs Teen Titans (DVD)
    \item Batman v Superman: Dawn of Justice
\end{itemize}
& X-Men-Apocalypse - The Cure/Come The Apocalypse VHS 
& \begin{itemize}[leftmargin=*, nosep, itemsep=-1pt, label=\textbullet]
    \item The Martian
    \item X-Men VHS
    \item The Young Riders: The Series - Season 1-3
    \item The Last Witch Hunter Digital
    \item The Martian: Extended Edition 4K Ultra-HD
    \item The Revenant
    \item The Hunger Games: Catching Fire 2013
    \item The Hobbit: The Desolation of Smaug
    \item The Hobbit: The Battle of the Five Armies [DVD] [2015]
    \item The Hobbit: An Unexpected Journey
\end{itemize}
& \Large \ding{55} \\ 
\cline{2-6}

& \multirow{11}{*}{\rotatebox{90}{After}}
& \begin{itemize}[leftmargin=*, nosep, itemsep=-1pt, label=\textbullet]
    \item The Hobbit: Battle of the Five Armies [DVD] [2015]
    \item Tinker Bell and the Neverbeast
    \item 55 Days at Peking (1900 Boxer Rebellion)
    \item The Incredibles (Mandarin Chinese Edition)
    \item The Legend of Longwood: A Magical Quest
    \item Batman vs. The Court of Owls
    \item The Last Witch Hunter
    \item Batman: Dark Knight Down
    \item Justice League vs Teen Titans: A Heroic Showdown
    \item Batman V Superman: Dawn of Justice (2016)
\end{itemize}
& X-Men: Apocalypse VHS (2017)
& \begin{itemize}[leftmargin=*, nosep, itemsep=-1pt, label=\textbullet]
    \item The Hobbit: An Unexpected Journey (2012)
    \item Captain America: Civil War (1)
    \item The Martian Stranded
    \item Guardians of the Galaxy (2014) [Region Free]
    \item \textbf{X-Men: Apocalypse VHS}
    \item Jurassic World
    \item Avengers: Ultron Rising
    \item Ant-man: The Tiny Hero
    \item The Revenant (Multi-Region PAL/NTSC DVD)
    \item The Big Short: A Tale of Financial Forecasts
\end{itemize}
& \Large \ding{51} \\
\hline

\multirow{22}{*}{\rotatebox{90}{\textbf{Case 2}}} 
& \multirow{11}{*}{\rotatebox{90}{Before}}
& \begin{itemize}[leftmargin=*, nosep, itemsep=-1pt, label=\textbullet]
    \item Barbie Fashion Complete Look 2-Pack, Pop Concert Set
    \item Barbie Complete Look Fashion Pack \#5
    \item Barbie Style Doll, Jean Jacket and Black/White Skirt
    \item Barbie Style Nikki Doll
    \item Barbie Fashion Complete Look 2-Pack, Movie Set
    \item Barbie Style Nikki Doll
    \item Barbie Fifth Harmony Lauren Doll
    \item Barbie Complete Look Fashion Pack \#3
    \item Barbie Style Summer Doll with Pink Paisley Dress and Jacket
    \item Barbie Complete Look Fashion Pack \#4
\end{itemize}
& Barbie Complete Look Fashion Pack \#8
& \begin{itemize}[leftmargin=*, nosep, itemsep=-1pt, label=\textbullet]
    \item Barbie Style Doll, Black and Silver Jacket
    \item Barbie Style Teresa Doll
    \item Barbie Style Midge Doll
    \item Barbie Style Raquelle Doll, Leopard Print Jacket
    \item Barbie Style Nikki Doll
    \item Barbie Style Doll, Jean Jacket and Black/White Skirt
    \item Barbie Style Glam Doll with Pink Retro Print Dress
    \item Barbie Style Barbie Doll
    \item Barbie Style Summer Doll with Pink Paisley Dress and Jacket
    \item Barbie Style Nikki Doll
\end{itemize}
& \Large \ding{55} \\
\cline{2-6}

& \multirow{11}{*}{\rotatebox{90}{After}}
& \begin{itemize}[leftmargin=*, nosep, itemsep=-1pt, label=\textbullet]
    \item Barbie Fashion 2-Pack: Pop Concert Outfits
    \item Barbie Fashion Pack \#5
    \item Barbie Fashion Doll with Jean Jacket and Skirt
    \item Barbie Fashion Doll (1)
    \item Barbie Fashion 2-Pack: Movie Date Outfits
    \item Barbie Style Nikki Doll: Fashionista Friend
    \item Barbie Fifth Harmony Lauren Doll
    \item Barbie Fashion Pack: 6 Dresses, 2 Coats, Shoes \& Purse
    \item Barbie Fashion Doll with Pink Dress and Jacket
    \item Barbie Fashion Pack \#4: 6 Dresses, 2 Coats, Shoes, and Purse
\end{itemize}
& Barbie Fashion Pack \#8
& \begin{itemize}[leftmargin=*, nosep, itemsep=-1pt, label=\textbullet]
    \item Barbie Fashion Doll with Dark Hair
    \item \textbf{Barbie Fashion Pack \#8}
    \item Barbie Fashion Pack \#6: 6 Dresses, 2 Coats, Shoes \& Purse
    \item Barbie Fashion Doll with Leather Jacket and Accessories
    \item Barbie Fashion Doll with Black and Silver Jacket
    \item Barbie Fashion Doll with Jean Jacket and Skirt
    \item Barbie Fashion Pack \#3: Dresses, Shoes, and Handbags
    \item Barbie Fashion Doll with Pink Dress and Jacket
    \item Barbie Fashion Doll with Accessories
    \item Barbie Fashion Pack \#5
\end{itemize}
& \Large \ding{51} \\
\hline
\end{tabular}%
}
\caption{Comparison of recommendation results. (\ding{51}) Success, (\ding{55}) Failure.}
\label{tab:case_study_multiline}
\end{table*}

The case study in Table~\ref{tab:case_study_multiline} illustrates how TR improves recommendation accuracy in different domains. In the movie domain, rewritten titles enhance semantic clarity, enabling the system to correctly rank relevant items—such as "X-Men: Apocalypse VHS"—within the top-10 recommendations, where previously it was missed. Similarly, in the toy domain, verbose and inconsistent original titles hindered accurate recommendations, but after rewriting, the system successfully identified and ranked items aligned with user preferences near the top. These examples demonstrate the practical impact of the TR module in enhancing recommendation relevance.

\section{Formal Definition of Out-of-domain (OOD) Recommendations}

For each dataset, let $\mathcal{I}$ denote the finite domain catalog of in-domain items. Given a user context $x$ (including interaction history and dialogue turns), a recommender $f_{\theta}$ produces a list of candidate recommendations
$R(x) = (\hat{i}_1(x), \dots, \hat{i}_T(x))$ in some output space (e.g., natural-language titles, item identifiers, or discrete codes). An individual prediction $\hat{i}_t(x)$ is \emph{out-of-domain} if it cannot be mapped to any catalog item, i.e., $\hat{i}_t(x) \notin \mathcal{I}$. The out-of-domain recommendation problem is to design models and grounding mechanisms such that, for all user contexts $x$ and all recommendation positions $t$, every generated item lies within the catalog, i.e., $\hat{i}_t(x) \in \mathcal{I}$.


\onecolumn
\begin{lstlisting}[caption=Prompts for training, label={lst:training_prompts}]
System: You are an expert recommender engine as well as a helpful, respectful and honest assistant.

Instruction 1: You need to generate a recommendation list considering user's preference from historical interactions.The historical interactions are provided as follows: {history}. You need to generate a recommendation list with {item_count} different items. Each item should be enclosed by <SOI> and <EOI>. <SOI> should be generated before item title, and <EOI> should be generated after item title.
Output: {item_list}

Instruction 2: You need to select a recommendation list considering user's preference from historical interactions.The historical interactions are provided as follows: {history}. The candidate items are: {candidate_titles}. You need to select a recommendation list with {item_count} different items from candidate items. Each item should be enclosed by <SOI> and <EOI>. <SOI> should be generated before item title, and <EOI> should be generated after item title.
Output: {item_list}

Instruction 3: Your task is generating a recommendation list according user's preference from historical interactions.The historical interactions are provided as follows: {history}. Please generate a recommendation list with {item_count} different items.
Output: {item_list}

Instruction 4: Your task is selecting a recommendation list according user's preference from historical interactions.The historical interactions are provided as follows: {history}. The candidate items are: {candidate_titles}. Please select a recommendation list with {item_count} different items from candidate items.
Output: {item_list}
\end{lstlisting}

\end{document}